\documentclass[12pt]{iopart}

\usepackage{graphics}
\usepackage{iopams}
\usepackage{color}
\usepackage{graphicx}
\usepackage{dcolumn} 
\usepackage{bm}
\usepackage{epsfig}
\usepackage{float}
\usepackage{textcomp}
\usepackage{booktabs}
\usepackage{sidecap}

\begin{document}

\title[$A$Fe$_{2}$As$_{2}$ - Theory and
  Experiment]{$A$Fe$_{2}$As$_{2}$ ($A$ = Ca, Sr, Ba, Eu) and
  SrFe$_{2-x}$$TM_{x}$As$_{2}$ ($TM$ = Mn, Co, Ni): crystal structure,
  charge doping, magnetism and superconductivity} \author{Deepa
  Kasinathan,$^{1}$ Alim Ormeci,$^{1}$ Katrin Koch,$^{1}$ Ulrich
  Burkhardt,$^{1}$ Walter Schnelle,$^{1}$ Andreas Leithe-Jasper,$^{1}$
  Helge Rosner$^{1}$} \address{$^1$Max-Planck-Institut f\"ur Chemische
  Physik fester Stoffe, Dresden, Germany }
\ead{Deepa.Kasinathan@cpfs.mpg.de, Rosner@cpfs.mpg.de}

\date{\today}

\begin{abstract}
The electronic structure and physical properties of the pnictide
compound families $RE$OFeAs ($RE$ = La, Ce, Pr, Nd, Sm),
$A$Fe$_{2}$As$_{2}$ ($A$ = Ca, Sr, Ba, Eu), LiFeAs and FeSe are quite
similar. Here, we focus on the members of the $A$Fe$_{2}$As$_{2}$
family whose sample composition, quality and single crystal growth are
better controllable compared to the other systems. Using first
principles band structure calculations we focus on understanding the
relationship between the crystal structure, charge doping and
magnetism in $A$Fe$_{2}$As$_{2}$ systems. We will elaborate on the
tetragonal to orthorhombic structural distortion along with the
associated magnetic order and anisotropy, influence of doping on the
$A$ site as well as on the Fe site, and the changes in the electronic
structure as a function of pressure.  Experimentally, we investigate
the substitution of Fe in SrFe$_{2-x}TM_{x}$As$_{2}$ by other 3$d$
transition metals, $TM$ = Mn, Co, Ni.  In contrast to a partial
substitution of Fe by Co or Ni (electron doping) a corresponding Mn
partial substitution does not lead to the supression of the
antiferromagnetic order or the appearance of superconductivity.  Most
calculated properties agree well with the measured properties, but
several of them are sensitive to the As $z$ position. For a
microscopic understanding of the electronic structure of this new
family of superconductors this structural feature related to the Fe-As
interplay is crucial, but its correct ab initio treatment still
remains an open question.
\end{abstract}

\submitto{\NJP}
\maketitle

\section{Introduction}
Physics community around the world has been tirelessly working for the
past months to find different ways of increasing the superconducting
transition temperature $T_{c}$ following the discovery of
superconductivity at 26 K in the rare-earth based ($RE$OFeAs) system
LaO$_{1-x}$F$_{x}$FeAs ($x$ = 0.05 - 0.12) \cite{Kamihara08a}.  Spirited
search by the experimentalists has led to eventually raising $T_{c}$
to 55 K for another member of this family of compounds,
SmFeAsO$_{0.9}$F$_{0.1}$ \cite{Ren08a}. Shortly afterwards, another
family of Fe-based compounds $A$Fe$_{2}$As$_{2}$ ($A$ = Ca, Sr, Ba,
Eu) was also found to be superconducting upon hole doping
\cite{definition} on the $A$ site with a maximum $T_{c}$ = 38 K
\cite{Sasmal08a,GFChen08a,Rotter08b}.  These discoveries are followed
by the announcements of other new parent compounds: LiFeAs, FeSe,
SrFeAsF with a maximum $T_{c}$ of 18 K \cite{WangXC08a}, 14 K
\cite{WangXC08a} (27 K using pressure \cite{Mizuguchi08a}) and 56 K
\cite{Wu08b}, respectively.  The basic features, that are common to
many of these new parent compounds, are the anti-ferromagnetic
ordering of the Fe-spins at $T_{N}$ $\approx$ 100--200 K, and the
quasi-2D nature of the electronic structure.  All the members
belonging to the above mentioned families have been shown to
superconduct upon either ``hole'' or ``electron doping'' only with the
exception of LiFeAs, wherein it is considered that the Li layer acts
as a charge reservoir for the system \cite{WangXC08a}.
Superconductivity also emerges upon application of hydrostatic
pressure for certain members of the $RE$OFeAs, $A$Fe$_{2}$As$_{2}$ and
FeSe families.  Notwithstanding the abundant research that has already
been performed, the microscopic nature of the mechanism of
superconductivity has thus far remained elusive.  A systematic study
of the many members within one particular family, using both
experimental and theoretical techniques, would pave way to a more
concrete understanding of the electronic structure of the normal
state. Although the $RE$OFeAs systems have larger $T_{c}$ values,
their sample preparation, characterization and quality, as is the case
for many oxides, have to be considered with great care. Alternatively
it has been demonstrated that the $A$Fe$_{2}$As$_{2}$ systems could be
synthesized comparatively easier.  Consequently, here we carry out a
systematic investigation using both theoretical and experimental
techniques for the $A$Fe$_{2}$As$_{2}$ systems.

All the $A$Fe$_{2}$As$_{2}$ compounds crystallize in the tetragonal
ThCr$_{2}$Si$_{2}$-type structure at room temperature.  They all
exhibit a structural transition upon cooling to an orthorhombic
lattice ($T_{0}$ for Ca $\approx$ 171 K \cite{Ronning08a}, Sr
$\approx$ 205 K \cite{Krellner08a}, Ba $\approx$ 140 K
\cite{Rotter08a}, Eu $\approx$ 200 K \cite{Jeevan08a}).  The
structural transition is coupled with an antiferromagnetic ordering of
the Fe moments with a wave vector $Q$ = [1,0,1] for the
spin-density-wave (SDW) pattern. Suitable substitution on either the
$A$ site or the Fe site can suppress the magnetic ordering, and then
the system becomes superconducting for certain ranges of doping (for
example, maximum $T_{c}$ = ~38 K (Ba,K)Fe$_{2}$As$_{2}$
\cite{Rotter08b}, ~32 K (Eu,K)Fe$_{2}$As$_{2}$ \cite{Jeevan08b}, ~21 K
Sr(Fe,Co)$_{2}$As$_{2}$ \cite{LeitheJasper08b},
Ba(Fe,Co)$_{2}$As$_{2}$ \cite{Sefat08b}).  Superconductivity can also
be induced in ``undoped'' and ``under-doped'' compounds by applying
pressure \cite{Kumar08a}.

In the following sections we describe the microscopic picture of the
magneto-structural transition and the effects of external pressure,
chemical pressure and charge doping on the $A$Fe$_{2}$As$_{2}$
systems.  In order to make the manuscript more easily readable, we
introduce the following abbreviations: $RE$OFeAs: $1111$,
$A$Fe$_{2}$As$_{2}$: $122$, LiFeAs: $111$, FeSe: $11$, SrFeAsF:
$1111^{\prime}$; ferromagnetic: FM, checkerboard (nearest neighbour)
antiferromagnetic: NN-AFM, columnar/stripe-type antiferromagnetic
order of the Fe-spins: SDW (spin-density-wave).

\section{Methods}

\subsection{{\bf Theory}}

We have performed density functional band structure calculations using
a full potential all-electron local orbital code FPLO
\cite{fplo1,fplo2,version} within the local (spin) density
approximation (L(S)DA) including spin-orbit coupling when needed.  The
Perdew-Wang \cite{perdew} parametrization of the exchange-correlation
potential is employed.  Density of states (DOS) and band structures
were calculated after converging the total energy on a dense $k$-mesh
with 24$\times$24$\times$24 points.  The strong Coulomb repulsion in
the Eu 4$f$ orbitals are treated on a mean field level using the
LSDA+$U$ approximation in the atomic-limit double counting scheme
\cite{czyzyk}.  Results we present below use the LSDA+$U$ method
\cite{aza} in the rotationally invariant form \cite{laz}.  In
accordance with the widespread belief that in the new Fe-based
superconducting compounds the Fe $3d$ electrons have a more itinerant
character than a localized one, and thereby are much less correlated
in comparison to the Cu $3d$ electrons in the high-$T_c$ cuprates, we
did not apply the LSDA+$U$ approximation to the Fe 3$d$ states.
Effects of doping on either the cation site or the Fe site were
studied using the virtual crystal approximation (VCA) treatment.  The
results obtained via VCA were cross checked using supercells for
certain doping concentrations.  The crystal structures are optimized
at different levels to investigate or isolate effects that may depend
sensitively to certain structural features.  The full relaxation of
the unit cell of the 122 systems at low temperatures involves
optimizing $a/b$ and $c/a$ ratios in addition to relaxing the As-$z$
position. 

\subsection{{\bf Experimental}}\label{experimental}

\begin{figure}[h]
  \begin{center}
    \begin{minipage}[t]{0.46\linewidth}
      \raisebox{-3cm}{\includegraphics[clip=, width=1.2\linewidth]{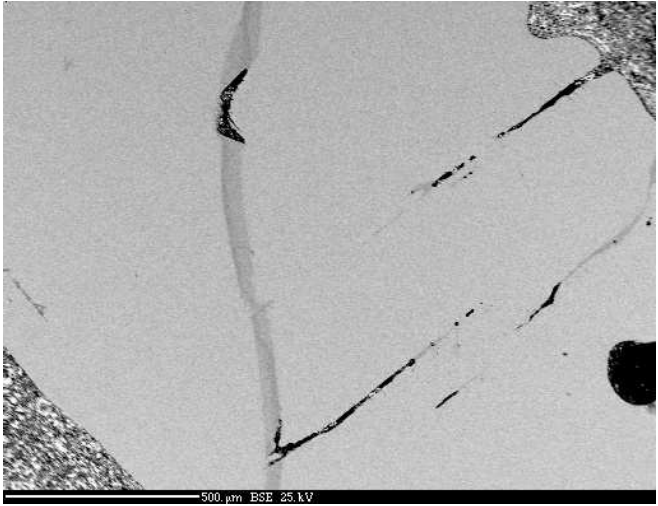}}
    \end{minipage}
    \begin{minipage}[t]{0.52\linewidth}
      \caption{\label{crystal} Micrograph of a polished crystal from the same 
       batch crystal X2 was selected. Light grey phase: bulk crystal. 
       Dark grey phase: inclusions of flux. Black regions: micro-cracks and
       cavities. }
    \end{minipage}
  \end{center}
\end{figure}

Polycrystalline samples were prepared by sintering in glassy-carbon
crucibles which were welded into tantalum containers and sealed into
evacuated quartz tubes for heat treatment at 900\textdegree C for 16
hours with two regrinding and compaction steps. First precursors SrAs,
Co$_{2}$As, Fe$_{2}$As, Mn$_{2}$As and NiAs were synthesized from
elemental powders sintered at 600\textdegree C for 48 h (Mn, Fe, Co,
Ni 99.9 wt.\%; As 99.999 wt.\%; Sr 99.99 wt.\%). These educts were
powdered, blended in stoichiometric ratios, compacted to pellets, and
heat treated. All sample handling was done inside argon-filled glove
boxes.  Crystals were grown in glassy-carbon crucibles by a modified
self flux method \cite{Morinaga2008,Sefat08b} in melts with
compositions SrFe$_{5-x}$Co$_{x}$As$_{5}$ (0.5 $\leq$ $x$ $\leq$ 0.85)
by cooling from 1250\textdegree C to 1100\textdegree C within 48
hours. The melt was spun off at 1100\textdegree C using a centrifuge
\cite{Bostrom2006}. Metallographic investigations were performed on
polished surfaces of selected secured crystal
platelets. Electron-probe microanalysis (EPMA) with
wavelength-dispersive analysis was accomplished in a Cameca SX100
machine.  Crystal X1 was grown in a flux
SrFe$_{4.25}$Co$_{0.75}$As$_{5}$ and crystal X2 in a flux
SrFe$_{0.15}$Co$_{0.85}$As$_{5}$.  For crystal X1 from EPMA the
composition (in at.\%)
Sr$_{19.9(2)}$Fe$_{36.2(1)}$Co$_{4.2(1)}$As$_{39.7(1)}$ was found
which corresponds to SrFe$_{2-x}$Co$_{x}$As$_{2}$ with $x$ $\approx$
0.21.  For crystal X2 from EPMA the composition (in at.\%)
Sr$_{19.5(2)}$Fe$_{35.5(1)}$Co$_{5.0(2)}$As$_{39.9(2)}$ was found
which corresponds to SrFe$_{2-x}$Co$_{x}$As$_{2}$ with $x$ $\approx$
0.25. Crystals grown from a flux SrFe$_{4.5}$Co$_{0.5}$As$_{5}$ had a
composition of Sr$_{19.9(3)}$Fe$_{36.9(1)}$Co$_{3.2(1)}$As$_{39.9(2)}$
corresponding to SrFe$_{2-x}$Co$_{x}$As$_{2}$ with $x \approx$ 0.15
and showed no superconductivity. All crystals grown up to now exhibit
some inclusions of flux-material which can be seen in
Fig.~\ref{crystal} (for crystal X2 the second phase has the
composition Sr$_{3(1)}$Fe$_{44(1)}$Co$_{8(1)}$As$_{45(1)}$ (in
at.\%)).

\section{Results - Theory}

\subsection{{\bf Ambient-temperature phase : tetragonal }}

\begin{figure}[H]
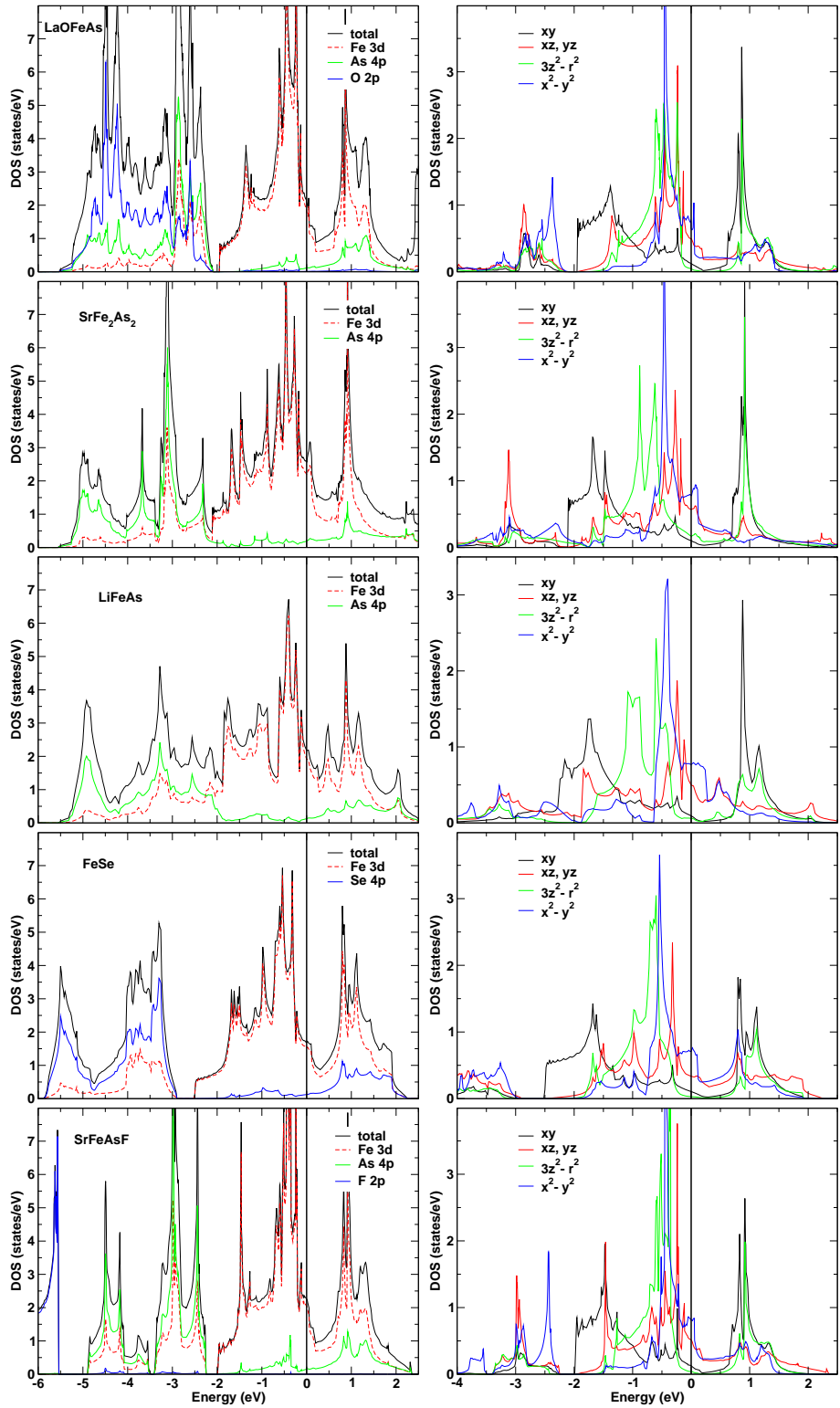

  \begin{center}
    \begin{minipage}[t]{0.38\linewidth}
      \epsfig{file=LaOFeAs_dos.eps, clip=, width=\linewidth}
    \end{minipage}
    \begin{minipage}[t]{0.38\linewidth}
      \epsfig{file=LaOFeAs_Fedos.eps, clip=, width=\linewidth}
    \end{minipage}
    \begin{minipage}[t]{0.38\linewidth}
      \epsfig{file=SrFe2As2_dos.eps, clip=, width=\linewidth}
    \end{minipage}
    \begin{minipage}[t]{0.38\linewidth}
      \epsfig{file=SrFe2As2_Fedos.eps, clip=, width=\linewidth}
    \end{minipage}
    \begin{minipage}[t]{0.38\linewidth}
      \epsfig{file=LiFeAs_dos.eps, clip=, width=\linewidth}
    \end{minipage}
    \begin{minipage}[t]{0.38\linewidth}
      \epsfig{file=LiFeAs_Fedos.eps, clip=, width=\linewidth}
    \end{minipage}
    \begin{minipage}[t]{0.38\linewidth}
      \epsfig{file=FeSe_dos.eps, clip=, width=\linewidth}
    \end{minipage}
    \begin{minipage}[t]{0.38\linewidth}
      \epsfig{file=FeSe_Fedos.eps, clip=, width=\linewidth}
    \end{minipage}
    \begin{minipage}[t]{0.38\linewidth}
      \epsfig{file=SrFeAsF_dos.eps, clip=, width=\linewidth}
    \end{minipage}
    \begin{minipage}[t]{0.38\linewidth}
      \epsfig{file=SrFeAsF_Fedos.eps, clip=, width=\linewidth}
    \end{minipage}
\caption{\label{dos}Comparison of total and site-resolved density of
  states (DOS) per cell (left panel) and Fe 3$d$ orbital resolved DOS
  (right panel) of a representative member for each of the new
  superconducting family of compounds. LaOFeAs:$1111$,
  SrFe$_{2}$As$_{2}$:$122$, LiFeAs:$111$, FeSe:$11$,
  SrFeAsF:$1111^{\prime}$. The solid vertical lines at zero energy
  denote the Fermi level $E_{F}$.}
  \end{center}
\end{figure}

We begin with comparing the DOS computed for a representative member
of each family of the iron pnictide compounds. In these calculations,
experimental values of the ambient-temperature tetragonal lattice
parameters and atomic positions were used for all the
systems. Collected in Fig.~\ref{dos} are the non-magnetic total and
orbital-resolved DOS for five systems: $1111$, $122$, $111$, $11$,
$1111^{\prime}$. The states close to the Fermi energy ($E_{F}$) in all
these systems are comprised mainly of Fe $3d$ contributions. The
contribution of the pnictide atom (or Se in the case of FeSe) to the
Fermi surface is small but non-zero.  A pseudo-gap-like feature in the
DOS slightly above the $E_{F}$ is common to all the systems.  At the
outset, all the five systems look quite similar to one another, but
slight differences are already visible when analyzing the Fe 3$d$
orbital resolved DOS, presented on right panels in Fig.~\ref{dos}. The
contribution of the Fe 3$d_{x^{2}-y^{2}}$ orbital (the orbital
pointing directly towards the nearest neighbour Fe ions) is the
largest close to $E_{F}$ for all the systems.  The corresponding bands
(not shown here) are highly dispersive in the $a-b$ plane and remain
flat along the $c$-axis, indicative of the quasi-2D nature of this
band.  The distance of the 3$d_{x^{2}-y^{2}}$ edge from the $E_{F}$
varies for the different systems and is the largest for the 111 family
and smallest for the 1111 family. The second largest contribution to
the Fermi surface comes from the doubly degenerate 3$d_{xz}$ and
3$d_{yz}$ orbitals and this feature is again consistent for all the
iron pnictide compounds.

\subsection{{\bf Structural distortion vs. magnetic order}}

\subsubsection{{\bf Tetragonal to orthorhombic distortion}}
\label{str}

As discussed above, the FeAs-based compounds crystallizing in
different structures have very similar electronic properties. However,
between the $1111$ and the $122$ families there is an important
difference in regard to structural and magnetic transitions. In the
former, the transition temperatures for the structural transition are
10--20 K higher than those of the magnetic one. On the other hand, for
the $122$ family the structural and the magnetic transitions are found
to be coupled and occur at the same temperature. By first-principles
calculations we explored the nature of this intimate connection
between the two transitions for the $122$ systems.

\begin{figure}[h]
  \begin{center}
    \begin{minipage}[t]{0.46\linewidth}
      \raisebox{-8cm}{\includegraphics[clip=, width=1.2\linewidth]{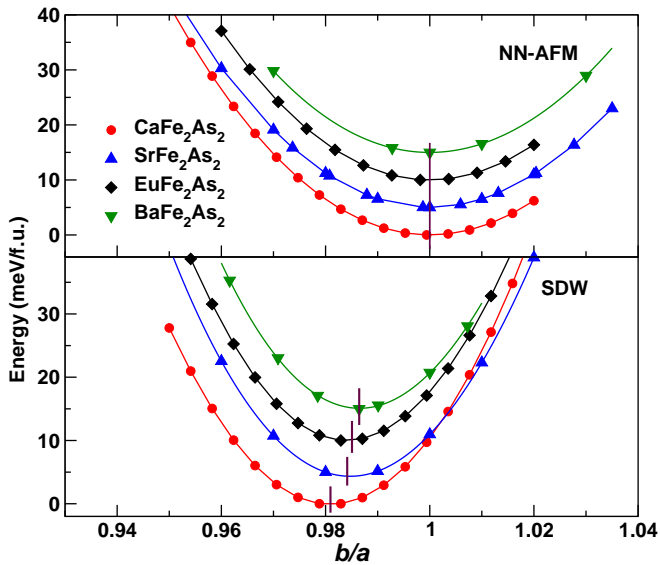}}
    \end{minipage}
    \begin{minipage}[t]{0.52\linewidth}
      \caption{\label{dist} Total energy as a function of $b/a$ for
        the $A$Fe$_{2}$As$_{2}$ ($A$ = Ca, Sr, Eu, Ba) systems with
        two possible spin arrangements between the Fe spins. In NN-AFM
        the spins of all the nearest neighbours are anti-parallel to
        each other. In the SDW pattern, the spins along the longer
        $a$-axis are anti-parallel while the spins along the shorter
        $b$-axis are parallel. The solid vertical lines denote the
        minimum value of the energy. The minimum energy for each
        system has been set to zero. For clarity, Sr122, Eu122 and
        Ba122 curves are offset by 5, 10 and 15 meV.}
    \end{minipage}
  \end{center}
\end{figure}

\begin{table}[htb]
\begin{center}
\caption{\label{bovera}Comparison of $b/a$ obtained from LSDA for the
  SDW pattern with experimental reports. The references from which the
  experimental numbers are obtained are indicated. With the exception
  of Ca122, (more detailed discussion, see text) the trend in the
  in-plane axis distortion for the orthorhombic structure is
  consistent with experimental findings.}
\begin{tabular}{cccc}
        & $b/a$-LSDA     &  $b/a$-expt    & Ref \\
\hline
Ca122   & 0.9809        &  0.9898       & \cite{Kreyssig08a}\\
Sr122   & 0.9841	 &  0.9889       & \cite{Jesche08a} \\
Eu122   & 0.9850	 &  0.9898       & \cite{Tegel08a}\\
Ba122   & 0.9864	 &  0.9928       & \cite{Rotter08a} \\
\hline
\end{tabular}
\end{center}
\end{table}
\noindent

We calculated total energies for different $b/a$ ratios for different
magnetic models. For the nonmagnetic (NM), ferromagnetic (FM) (not
shown) as well as the nearest-neighbour antiferromagnetic (NN-AFM)
(Fig.~\ref{dist}, upper panel) patterns, lowest total energy occurred
at $a = b$, indicating that for these patterns the tetragonal
structure is more stable than the orthorhombic structure. Inclusion of
spin-orbit effects did not change this result. The tetragonal to
orthorhombic distortion is obtained only for the SDW pattern as
displayed in the lower panel of Fig.~\ref{dist}. Hence, LSDA
calculations clearly show that the SDW state is necessary for the
tetragonal-to-orthorhombic transition to take place. The size of this
effect, namely the deviation of the calculated $b/a$ ratio from unity,
depends on the size of the cation, the ratio being smallest in
BaFe$_2$As$_2$, and largest in CaFe$_2$As$_2$.  The values obtained
via LSDA are collected in Table~\ref{bovera} and compared to the
experimental reports. The trend obtained in LSDA fits well with the
experimental $b/a$ ratios with the exception of CaFe$_{2}$As$_{2}$.
Experimentally, CaFe$_{2}$As$_{2}$ crystals are shown to have complex
microstucture properties. Recent studies using transmission electron
microscopy (TEM) \cite{Ma2008} have shown a pseudo-periodic modulation
and structural twinning arising from tetragonal to orthorhombic
transition only in CaFe$_{2}$As$_{2}$ but not in Sr or Ba $122$
systems. A structural twinning hinders the correct estimation of the
lattice parameters and thereby may explain the experimental deviation
in the trend of $b/a$ ratio for CaFe$_{2}$As$_{2}$ with respect to
LSDA.

In all four compounds the coupling along the shorter in-plane axis is
FM in agreement with experimental findings. Although these results are
robust with respect to details of structure and calculations, the
preferred direction of the spins are found to be quite sensitive. We
performed fully-relativistic calculations for $A =$ Ca, Sr and Ba
cases using (i) SDW with $Q =$ [1 0 0], (ii) SDW with $Q =$ [1 0 1].
The latter SDW pattern requires the doubling of the $c$ lattice
parameter and the corresponding calculations are highly time
consuming.  Therefore, only the cartesian axes are considered for
possible spin orientations, and the structural data corresponding to
the minima in Fig.~\ref{dist} are used.  In the Sr122 case for both
SDW patterns we find the direction of the AFM coupling (along the
longer $a$-axis) as the easy axis in agreement with the neutron
scattering study result \cite{Jesche08a}.  However, in Ca and Ba 122
cases different axes in the $(a,b)$ plane are found as easy axes for
different SDW patterns. The longer $a$-axis, is the easy axis for Ba
(Ca) 122 for $Q =$ [1 0 0] ($Q =$ [1 0 1]); the shorter axis, $b$, the
direction of FM coupling, for Ba (Ca) 122 for $Q =$ [1 0 1] ($Q =$ [1
  0 0]).  Since the involved energy differences are tiny (of the order
of 15--30 $\mu$eV per atom), a satisfactory resolution of this issue
requires further study.  Experimental study on
Ba122 \cite{Huang08} has been able to determine that spins lie along
the longer $a$-axis.

\subsubsection{{\bf Plasma frequency and effective dimensionality}}

Band structure calculations can provide information on the ``effective
dimensionality'' in a compound through various means, such as
dispersionless (flat) energy bands along certain symmetry lines, van
Hove singularities in DOS, etc. A simple quantitative measure,
however, can be obtained by computing plasma frequencies along the
main unit cell axes. For all of the $122$ compounds as well as the
representative compounds for the other families, plasma frequencies
are calculated by nonmagnetic calculations using the experimental
structural data of the tetragonal phase.  In Table~\ref{plasma} we
present the ratio of the in-plane plasma frequency
$(\omega_{\mathrm{P}}^a = \omega_{\mathrm{P}}^b)$ to the plasma
frequency along the $c$-axis $(\omega_{\mathrm{P}}^c)$. One notices
that for the $122$ family the results are in line with expectations:
the compound is least anisotropic (more 3D-like) for the smallest
cation (Ca), and strongly anisotropic (less 3D-like) for the largest
cation (Ba). Additionally, the $1111$ and $1111^{\prime}$ families are
seen to be much more anisotropic (more 2D-like) than all of the
others.

\begin{table}[t]
\begin{center}
\caption{\label{plasma}Ratio of the in-plane plasma frequency
  $\omega_{\mathrm{P}}^{a}$ to the out-of-plane plasma frequency
  $\omega_{\mathrm{P}}^{c}$ for various members of the Fe-based
  superconducting systems. The observed trend in the plasma frequency
  ratios follows the trend in the ratios of the (inter-layer to
  intra-layer) Fe-Fe distances,
  $d_{c}^{\mathrm{Fe-Fe}}$/$d_{a}^{\mathrm{Fe-Fe}}$. As we go down the
  column, the systems go from being more 2D towards being more 3D. The
  maximum superconducting transition temperature
  $T_{c}^{\mathrm{max}}$ obtained either via doping or pressure is
  also collected in the last column. The trend in
  $T_{c}^{\mathrm{max}}$ also follows the trend in
  $\omega_{\mathrm{P}}^{a}/\omega_{\mathrm{P}}^{c}$, with decreasing
  temperatures when the systems become more 3D. }
\begin{tabular}{ccccc}
                  &$\omega_{\mathrm{P}}^{a}/\omega_{\mathrm{P}}^{c}$& $c/a$ &  $d_{c}^{\mathrm{Fe-Fe}}$/$d_{a}^{\mathrm{Fe-Fe}}$ & $T_{c}^{\mathrm{max}}$ (K)\\
\hline
SrFeAsF           &     19.892                &2.2426 &  3.1715  & 56 \cite{Wu08b} \\
LaOFeAs           & 	8.9467		      &2.1656 &	 3.0626  & 55 \cite{Ren08a} \\
FeSe              &	4.1119                &1.4656 &  2.0727  & 27 \cite{Mizuguchi08a} \\
LiFeAs            &   	3.2181                &1.6785 &	 2.3738	 & 18 \cite{WangXC08a} \\
BaFe$_{2}$As$_{2}$&   	3.2926	              &3.2850 &  2.3228  & 38 \cite{Rotter08b} \\
SrFe$_{2}$As$_{2}$&   	2.8329	              &3.1507 &  2.2279  & 38 \cite{GFChen08a,Sasmal08a} \\
CaFe$_{2}$As$_{2}$&	1.3953                &3.0287 &  2.1416  & 20 \cite{Wu08a} \\
\hline
\end{tabular}
\end{center}
\end{table}

In the case of iron arsenides, in a rather simplified picture, one
expects the plasma frequency ratio to be proportional to the ratio of
the shortest interlayer $d_c^{\mathrm{Fe-Fe}})$ to the shortest
intralayer Fe-Fe distances $(d_a^{\mathrm{Fe-Fe}})$. In terms of
lattice parameters the distance ratio
$d_c^{\mathrm{Fe-Fe}}/d_a^{\mathrm{Fe-Fe}}$ is $c/\sqrt2 a$ for
$122$'s and $\sqrt2 c/a$ for the others. Table~\ref{plasma} shows that
these two ratios are largely correlated with apparently the exception
of FeSe \cite{footnote3}.

\subsubsection{{\bf Effect of the As $z$ position and magnetic moments from
 LDA:}}
\label{Aszpos}

In any first-principles study of a magnetic system, an essential
aspect is the comparison of the computed magnetic moments with the
experimentally deduced ones.  This standard procedure proves to be
quite tricky in these FeAs-based compounds due to the unexpected
sensitivity of the magnetism to the As $z$ position \cite{Mazin08a}.
We performed a series of calculations \cite{footnote1} for Ca, Sr and
Ba 122 systems for FM, NN-AFM and SDW spin patterns using the
experimental volume (ambient pressure, below $T_{N}$) both with $a =
b$ and $a \neq b$ and optimizing the Fe-As distance (via As $z$
position) for each case.  Table~\ref{mom} compares the Fe moments
computed for the SDW pattern (the lowest-total-energy spin pattern
among those considered) with the experimental values obtained using
neutron diffraction and muon spin rotation ($\mu$SR).  In comparison
to the situation in the 1111 systems, here for the 122 systems, the
agreement between theory and experiment is seen to be better.
However, the computed Fe magnetic moment is found to increase from Ca
to Ba 122, whereas the trend is just the opposite according to the
estimation of moments from $\mu$SR results. The moments obtained from
neutron diffraction experiments are more reliable and remain rather
constant for the three $122$ systems considered here.  The influence
of the orthorhombic distortion on the calculated moments is quite
negligible.

The better agreement between theory and experiment regarding the Fe
magnetic moment in the case of 122 systems can be understood as
follows.  The computed \textit{vs} measured magnetic moment
discrepancy in the 1111 systems is usually explained to be a result of
large spin fluctuations \cite{Mazin08b}.  It is also known that spin
fluctuation effects are reduced when going from 2D systems to 3D
systems.  Hence, the results presented in Table~\ref{mom} provide
additional support for the effective dimensionality considerations
described above: the 122 systems have a more pronounced 3D nature than
the 1111 systems.  Since the description of spin fluctuation effects
is insufficient in LSDA, the computed values are immune to such
effects, while, of course, the values deduced from experiments do
reflect these effects.  Furthermore, since the Ba 122 system is more
2D-like than the Ca 122 system (\textit{cf.}  Table~\ref{plasma}), it
is expected to exhibit a smaller Fe moment (stronger spin fluctuation
effects), and this is in agreement with the $\mu$SR results.

\begin{table}[htb]
\begin{center}
\caption{\label{mom}Comparison of the magnetic moments in
  $\mu_{\mathrm{B}}$ per atom calculated using LSDA with the
  experimental values obtained via $\mu$SR (a local probe) and neutron
  diffraction measurements. In LSDA, the moments were calculated for
  the SDW pattern for the Fe spins using a tetragonal lattice ( $a =
  b$ ) and orthorhombic lattice ( $a \neq b$ ). The Fe-As distance has
  been optimized for all the calculations reported in this table.}
\begin{tabular}{ccccc}
            &  \multicolumn{2}{c}{LDA - SDW} &   $\mu$SR    & Neutron  \\
\cline{2-3}
            &   $a = b$  &  $a \neq b$ &                  & diffraction \\
\hline
Ca122       & 0.818    &   0.875     & 0.9 \cite{TGoko08a}   & 0.80 \cite{Goldman08a} \\
Sr122       & 1.1      &   1.13      & 0.8 \cite{TGoko08a}   & 0.94 \cite{Zhao08a} \\
Ba122       & 1.12     &   1.17      & 0.5 \cite{TGoko08a}   & 0.87 \cite{Matan08a} \\
\hline
\end{tabular}
\end{center}
\end{table}

\begin{figure}[htb]
\begin{center}\includegraphics[%
  clip,
  width=8.5cm,
  angle=0]{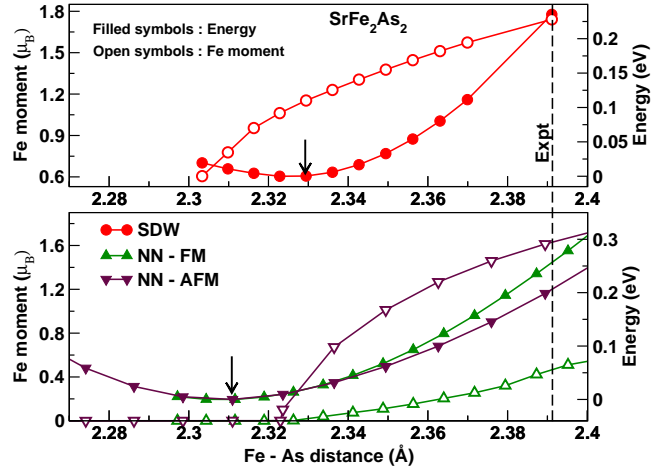}\end{center}
\caption{\label{Asz} Energy and Fe magnetic moment as a function of
  the Fe-As distance using different spin patterns for
  SrFe$_{2}$As$_{2}$ at the experimental volume around 90 K
  \cite{Tegel08a}. Optimization using FM and NN-AFM pattern leads to a
  nonmagnetic solution, while the SDW pattern stabilizes with a Fe
  moment of 1.13 $\mu_{B}$. The energy curves have been shifted by
  setting the minimum energy value to zero. The dashed vertical line
  refers to the experimental Fe-As distance obtained from
  Ref. \cite{Tegel08a}. The arrows indicate the position of the energy
  minima.  }
\end{figure}

The interplay between the Fe-As distance and different magnetic
long-range orders is illustrated in Fig.~\ref{Asz} for
SrFe$_{2}$As$_{2}$.  Similar results are found for the other $122$
systems.  The experimental value of the Fe-As distance in
SrFe$_{2}$As$_{2}$ in the orthorhombic phase is reported to be 2.391
\AA\ at 90 K \cite{Tegel08a}.  Using the FM or the NN-AFM spin pattern
produces a minimum in energy at around 2.31 \AA\ along with a complete
loss of the Fe magnetic moment.  In the SDW pattern, the optimum value
of the Fe-As distance is 2.327 \AA, only slightly larger than that
obtained using the FM or NN-AFM pattern, but the Fe magnetic moment is
still 1.13 $\mu_{\mathrm{B}}$.

The As-$z$ position is most likely one of the key issues for
understanding of the iron pnictides.  Present day density functional
theory based calculations using LDA (described above) and also GGA
\cite{Mazin08a,Yildirim08a} are not able to reproduce all the
experimental findings consistently. Since the magnetism and therefore
the superconductivity crucially depends on this structural feature and
the related accurate description of the Fe-As interaction, the
improvement of the calculations in this respect may offer the key to
the understanding of superconductivity in the whole family.

\subsection{{\bf Effects of pressure}}
\label{pressure}

\begin{figure}[t]
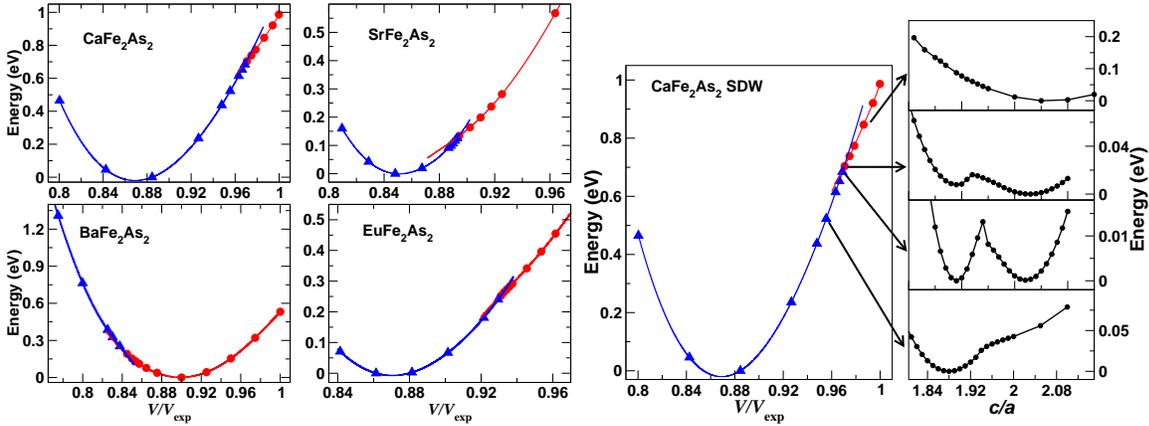

  \begin{center}
    \begin{minipage}[t]{0.48\linewidth}
      \epsfig{file=EvsVcomp.eps, clip=, width=\linewidth}
    \end{minipage}
    \begin{minipage}[t]{0.48\linewidth}
      \epsfig{file=EvsVwithca.eps, clip=, width=\linewidth}
    \end{minipage}
\caption{\label{EvsV}{\bf Left:} Energy as a function of volume for
  $A$Fe$_{2}$As$_{2}$ systems, with $c/a$ ratio optimized. The two
  curves correspond to the antiferromagnetically ordered Fe spins in
  the SDW pattern with a non-zero moment (red circles) and zero moment
  (blue triangles). The kink at the intersection of these two curves
  is caused by a collapse of the $c/a$ ratio upon pressure, which
  happens at the juncture when the systems lose their Fe moments and
  become non-magnetic. The $c/a$ ratio collapse is most pronounced in
  CaFe$_{2}$As$_{2}$ due to the small size of the Ca ion and as the
  size of the $A$ ions increases, this feature becomes more and more
  subtle. {\bf Right:} The evolution of the $c/a$ ratio collapse for
  CaFe$_{2}$As$_{2}$. Notice the emergence of the double minima.  }
  \end{center}
\end{figure}

\begin{figure}[h]
  \begin{center}
    \begin{minipage}[t]{0.46\linewidth}
      \raisebox{1cm}{\includegraphics[clip=, width=\linewidth, angle=-90]{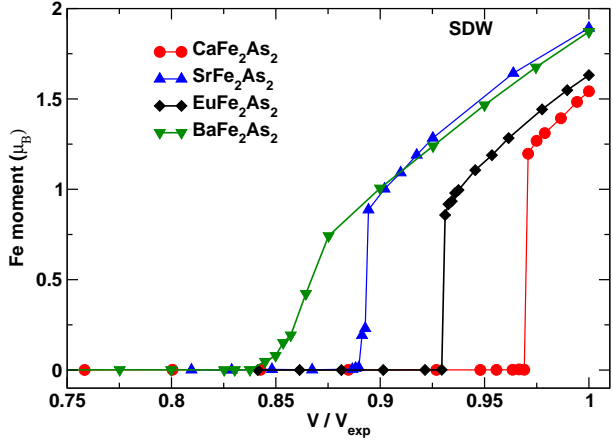}}
    \end{minipage}
    \begin{minipage}[t]{0.52\linewidth}
      \caption{\label{momvsV} Volume dependence of the Fe moments
        corresponding to the energy-volume curves in
        Fig.~\ref{EvsV}. The quenching of the spin magnetic moment is
        highly abrupt for the compounds with pronounced $c/a$ ratio
        collapse. The Fe moment in BaFe$_{2}$As$_{2}$ goes to zero
        smoothly reflecting the much smoother $c/a$ variation obtained
        for this compound. }
    \end{minipage}
  \end{center}
\end{figure}

\begin{figure}[b]
  \begin{center}
    \begin{minipage}[t]{0.48\linewidth}
      \raisebox{-7cm}{\includegraphics[clip=, width=1.3\linewidth, angle=0]{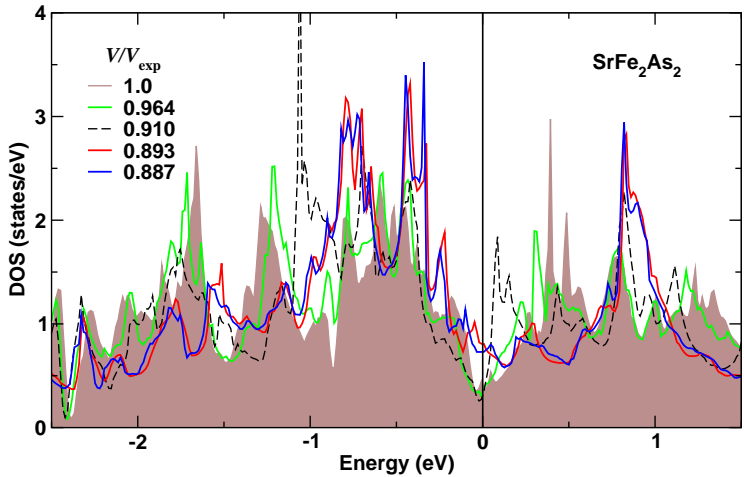}}
    \end{minipage}
    \begin{minipage}[t]{0.5\linewidth}
      \caption{\label{dospressure} Total DOS per cell as a function of
        reduced volume for SrFe$_{2}$As$_{2}$ corresponding to the
        energy-volume curves in Fig.~\ref{EvsV}. In the SDW pattern,
        the total DOS for both the spin-up and spin-down channels are
        the same. Therefore, we show here the DOS from only one of the
        spin channels. The DOS at the Fermi level $E_{F}$ at first
        decreases with decreasing volume, but later starts to increase
        after the Fe spin magnetic moment is quenched. }
    \end{minipage}
  \end{center}
\end{figure}

As mentioned above, superconductivity in the FeAs based systems can be
achieved via doping of charge carriers.  This kind of chemical
substitution is quite convenient, but changes the electronic structure
of the doped systems in a non-trivial way as compared to the undoped
systems. Using external pressure as a probe on these systems creates a
similar effect as doping, without the added complexity.  Application
of hydrostatic pressure suppresses both the tetragonal to orthorhombic
distortion and the formation of the SDW, and leads to the onset of
superconductivity in a similar fashion as charge doping.  All the
parent members of the $122$ family have been reported to superconduct
or show signs of its onset under pressure
\cite{Torikachvili08a,Kreyssig08a,Kumar08a,Miclea08a,Alireza08a}.
Ca122 was originally reported to superconduct ($T_{c}$ $\approx$ 10 K)
at 0.4 GPa pressure \cite{Torikachvili08a,Kreyssig08a} but recent work
by Yu and collaborators \cite{Yu08a} do not observe any signature of
bulk superconductivity and suggest a possible phase separation due to
non-hydrostatic conditions. Similarly, Sr122 was first reported to
superconduct at 2.8 GPa pressure with a $T_{c}$ of 27 K
\cite{Alireza08a}, while on the contrary Kumar {\it et al.}
\cite{Kumar08a} did not observe a zero-resistance state up to 3 GPa
pressure. A sharp drop in resistivity above 2 GPa pressure was
reported for Eu122, suggesting the onset of superconductivity and,
more interestingly, signatures of possible re-entrant
superconductivity.  Though there exist conflicting reports of the
transition pressure and possible phase separation in the sample under
pressure, it is worthwhile to explore the evolution of the electronic
structure and magnetism as a function of pressure.  Pressure studies
\cite{Kreyssig08a} on the CaFe$_{2}$As$_{2}$ system report a
significant $c/a$ collapse along with a structural transition
(orthorhombic to tetragonal) under modest pressures of less than 0.4
GPa, while no such collapse has been reported for the other compounds.
X-ray diffraction refinements carried out at 180 K for
SrFe$_{2}$As$_{2}$ \cite{Kumar08a} observe an orthorhombic to
tetragonal transition above 3.8 GPa, but no collapse of the $c/a$
ratio is observed.  Band structure calculations allow for the study of
such features up to very large pressures, that might not be easily
attainable through experiments.  We have calculated energy as a
function of volume for all the four systems in the $122$
family. Firstly, we wanted to investigate the possibility of a $c/a$
collapse for each member of the $122$ family.  Therefore, we
calculated energy as a function of volume using the SDW pattern and
optimizing only the $c/a$ ratio at each volume. The internal As-$z$
parameter was kept fixed at the experimental (room temperature)
value. The results from these calculations are collected in
Fig.~\ref{EvsV}.  Surprisingly, all the four $122$ systems have a kink
in the energy-volume curve caused by a non-continous change in the
$c/a$ ratio, which happens at the juncture when the systems lose
their Fe moments and become non-magnetic (see Fig.~\ref{momvsV}).  The
sharpness of the kink is largest for Ca122 and decreases with the
increasing size of the $A$ ion. Such an $A$-ion size effect was also
observed and discussed in section 3.2.1 when investigating the size of
the orthorhombic $b/a$ ratio.  Our result obtained from LSDA (via a
common tangent construction) for CaFe$_{2}$As$_{2}$ is in excellent
agreement with the previously reported \cite{Kreyssig08a} experimental
data, volume collapse: $\delta V^{\mathrm{LDA}}$ $\approx$ 4.7\%,
$\delta V^{\mathrm{exp}}$ $\approx$ 5\%; ratio collapse:
$\delta$($c/a$)$^{\mathrm{LDA}}$ $\approx$ 9.8\%,
$\delta$($c/a$)$^{\mathrm{exp}}$ $\approx$ 9.5\%. Experimentally
\cite{Kreyssig08a} a pressure of $\sim$ 0.3 GPa induces a transition
from the orthorhombic phase to a collapsed non-magnetic tetragonal
phase for CaFe$_{2}$As$_{2}$.  Scaling the volumes of the different
$122$ systems with respect to the experimental values
($V/V_{\mathrm{exp}}$), one observes that the kink in the
energy-volume curve for the other three $122$ systems happens at lower
volume ratios (or larger pressures) as compared to the Ca122. It
should be worthwhile to investigate this structural feature
experimentally by applying higher pressures to the Sr, Ba and Eu $122$
systems.

Changes in the electronic structure as a function of reduced volumes
were carefully monitored. Shown in Fig.~\ref{dospressure} are the
total DOS for SrFe$_{2}$As$_{2}$ at selected volumes. Similar results
are obtained for other $122$ systems. The DOS at $E_{F}$ decreases
gradually at first for up to 10\% volume reduction with respect to the
experimental volume. The Fe ions continue to carry a magnetic moment
though the actual values are quite reduced. Upon further reduction of
the volume, the spin moments get quenched and the DOS at $E_{F}$ begin
to increase and the system becomes non-magnetic. At the experimental
volume, the net moments on the various Fe orbitals are quite similar;
with the Fe-3$d_{x^{2}-y^{2}}$ having a slightly larger value than the
other orbitals. With the application of pressure, the net moments of
all the five $d$ orbitals decrease in a similar fashion and tend to
zero.

\begin{figure}[t]
\begin{center}\includegraphics[%
  clip,
  width=10cm,
  angle=0]{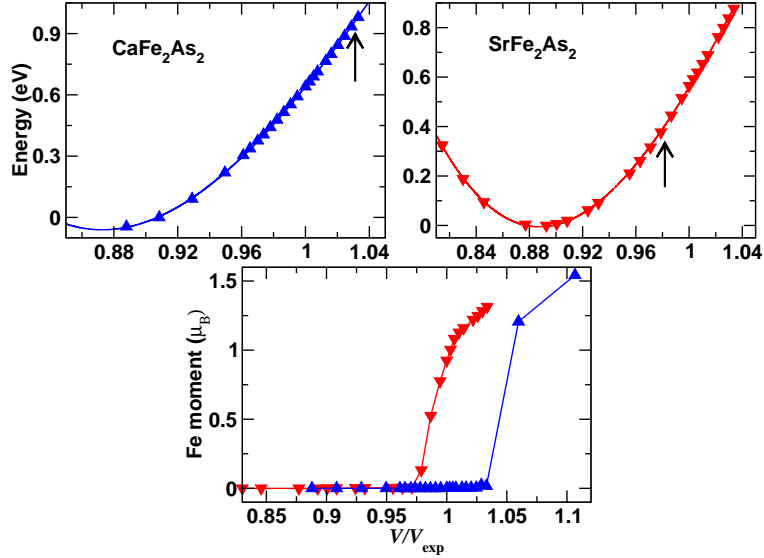}\end{center}
\caption{\label{fullopt} {\bf Top panel:} Energy as a function of
  volume after a full relaxation of all the parameters for
  SrFe$_{2}$As$_{2}$ and CaFe$_{2}$As$_{2}$. Contrary to the results
  depicted in Fig.~\ref{EvsV}, we now no longer observe any kink at
  the juncture when the Fe ions lose the magnetic moment
  (corresponding data point is indicated using an arrow). {\bf Bottom
    panel:} Fe magnetic moment as a function of volume for the Ca and
  Sr 122 systems. The results shown here are different from the moment
  values collected in Table.~\ref{mom}, because no $c/a$ optimization
  was carried out for the latter.  }
\end{figure} 

Another important feature that needs to be addressed in regard to the
energy-volume curves is the serious underestimation of the equilibrium
volume within LDA.  Generally, equilibrium volumes obtained from LDA
are smaller within up to 8\% of the experimentally reported values. In
the case of 122 systems using SDW pattern, we obtain values that are
13\%, 15\%, 10\%, and 13\% smaller than the experimental reports for
Ca, Sr, Ba and Eu 122 systems respectively. Moreover, at the LSDA
equilibrium volume, contrary to the experimental reports, all the
parent compounds are computed to be nonmagnetic.  The reason for this
discrepancy is unclear.  In the previous sections we discussed the
pronounced sensitivity of the Fe moments to the various structural
parameters in these FeAs systems.  Although with partial optimization
(fixed As-$z$ parameter) we have satisfactorily accounted for the
experimentally observed $c/a$ collapse in CaFe$_{2}$As$_{2}$, it is
necessary to find out what the ultimate LSDA solution is regarding the
geometrical structure and magnetism in the 122 systems. Consequently,
for Ca and Sr 122 systems and using the SDW pattern, at each volume we
have optimized all three free structural parameters in the following
order: 1) As-$z$ position, 2) $c/a$ ratio, 3) $b/a$ ratio.  This
sequence of steps has been repeated until the energies obtained are
converged to an accuracy of 10$^{-6}$ eV.  Collected in
Fig.~\ref{fullopt} are the energy-volume curves for the Ca and Sr
$122$ systems.  The equilibrium volume obtained after a full
optimization is only slightly larger than the values obtained after
just a $c/a$ optimization (see Fig.~\ref{EvsV}).  Contrary to the
results depicted in Fig.~\ref{EvsV}, we now no longer observe any kink
at the juncture when the Fe ions lose their magnetic moments. The
loss of moment for Sr122 is more gradual than for Ca122.  As volume is
decreased (higher pressures are applied) the Fe spin magnetic moments
tend to zero while the orthorhombic distortion ratio $b/a$ tends to
unity so that at increased pressures the tetragonal lattice is
favored. This observation of the lattice structure preferring the
tetragonal symmetry when the Fe ions become nonmagnetic is consistent
with our previous results in Sec.~\ref{str}, and reaffirms the
intimate connection between structure and magnetism for the 122
systems.  Optimizing the As $z$ parameter again tends to confirm
certain experimental findings (the connection between SDW magnetic
pattern and orthorhombic distortion) but not all (for example, lack of
$c/a$ collapse under pressure for CaFe$_{2}$As$_{2}$).  This again
re-affirms the need for a correct description of the Fe-As interplay
to obtain consistent results.

\subsection{{\bf Effects of charge doping} }

\begin{figure}[htb]
\begin{center}\includegraphics[%
  clip,
  width=8.5cm,
  angle=0]{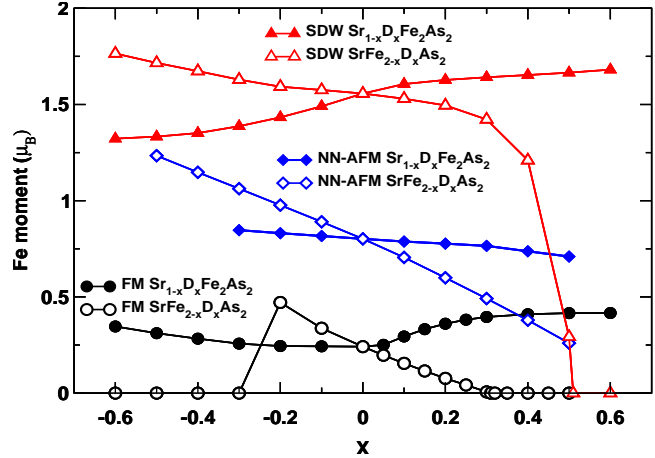}\end{center}
\caption{\label{vca} Results from the VCA calculations. Magnitude of
  the Fe moment as a function of charge doping (both on Sr site as
  well as Fe site) for different ordering patterns in
  SrFe$_{2}$As$_{2}$.  When {\bf $x$} is positive: electron doping;
  when {\bf $x$} is negative: hole doping.  The filled symbols and the
  open symbols indicate doping of the $A$ site and the Fe site
  respectively. `D' represents the dopant. Very different effects are
  observed when doping charge carriers on the Sr or Fe site. Magnetism
  is weakened when electrons are substituted on the Fe site, while
  strengthened when electrons are substituted on the Sr site. The
  relative trend between the different ordering patterns remains the
  same.  We have used the experimental lattice parameters (at
  $\approx$ 300 K) and As-$z$ value for all the calculations. }
\end{figure}

\begin{figure}[htb]
  \begin{center}
    \begin{minipage}[t]{0.48\linewidth}
      \epsfig{file=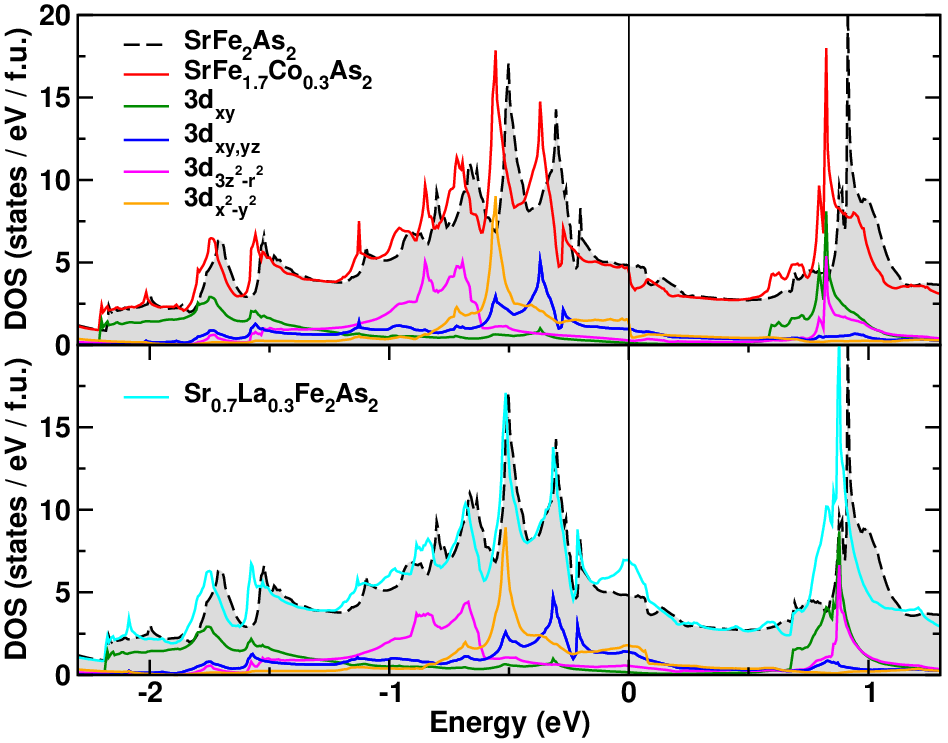, clip=, width=\linewidth}
    \end{minipage}
    \begin{minipage}[t]{0.48\linewidth}
      \epsfig{file=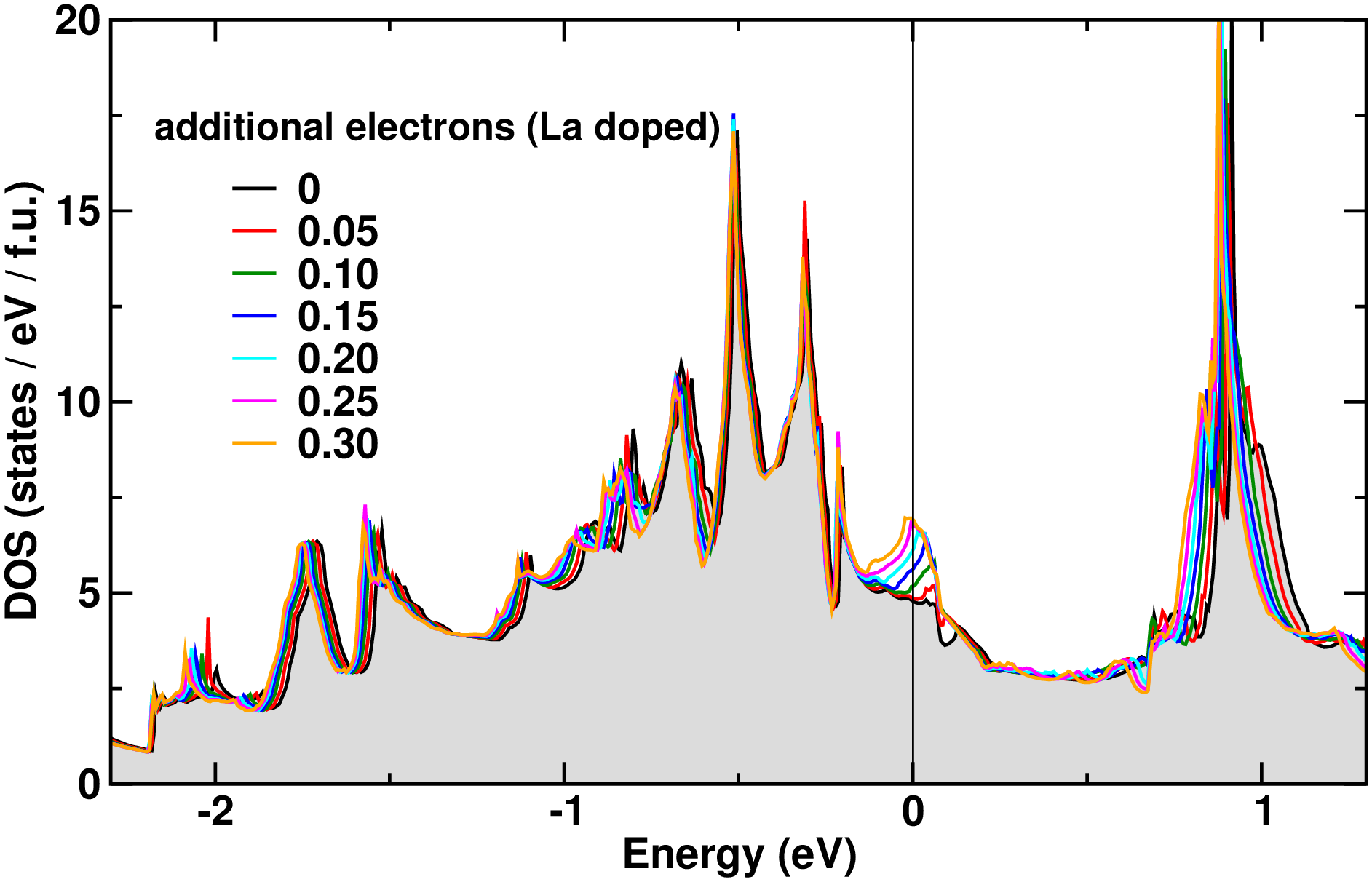, clip=, width=\linewidth}
    \end{minipage}
\caption{\label{covca}{\bf Top Left:} Non magnetic total and Fe
  orbital resolved DOS from a VCA calculation for a 15\% electron
  doping on the Fe site in Sr122. Upon electron doping, the DOS
  remains unchanged and displays a rigid-band-like behaviour.  {\bf
    Bottom Left:} Non-magnetic total and Fe orbital resolved DOS for a
  30\% electron doping on the Sr site in Sr122. Addition of electrons
  changes the shape of the DOS close to $E_{F}$ drastically. A
  pronounced peak starts to appear close to $E_{F}$ which tends to
  destabilize the system. The changes in the total DOS for various
  doping concentrations are shown in the right panel.  {\bf Right:}
  Non magnetic total DOS as a function of additional electrons on the
  Sr site in Sr122.  }
  \end{center}
\end{figure}

In order to understand the influence of charge doping (both electrons
and holes) on the electronic structure and henceforth the magnetism,
we performed total energy calculations using the VCA for three
different spin patterns: FM, NN-AFM and SDW.  We considered both
electron and hole doping on the $A$ site as well as the Fe site.  Our
results for the changes in the Fe moment in SrFe$_{2}$As$_{2}$ are
collected in Fig.~\ref{vca}. Similar results were obtained for other
members of the 122 family. We observe very different effects depending
on the sign (electrons or holes) and site (Sr or Fe) of the doping.
Regardless of the choice of magnetic ordering, electron doping on the
Fe (Sr) site weakens (enhances) the magnetism.  This behaviour can be
explained by analyzing how the nonmagnetic electronic structure
changes with electron doping (see Fig.~\ref{covca}).  Electron doping
on the Fe site (left upper panel of Fig.~\ref{covca}) results in DOS
very similar to that of the undoped case, main effect being the
$E_{F}$ moved toward higher energies to accommodate the added
electrons.  On the other hand, electron doping on the Sr site changes
the resultant DOS drastically (see left lower and right panel of
Fig.~\ref{covca}). Most of the major changes to DOS occur in the close
vicinity of $E_{F}$ giving rise to pronounced peaks at the $E_{F}$.  A
large value of DOS at the Fermi level, $N(E_F)$, is usually a sign of
instability for an electronic system.  The system can lower $N(E_F)$
by, for example, developing a long-range magnetic order provided the
Stoner criterion is satisfied.  The larger values of the computed Fe
magnetic moments may reflect such an increased instability to magnetic
order.  Additionally, with reservations for possible thermodynamical
considerations, appearance of this feature may explain why La doped
122 samples could not be synthesized until now.

Substitution of holes on the Sr site does not introduce significant
changes to the Fe magnetic moment.  Substitution of holes on the Fe
site tends to enhance magnetism for both AFM patterns whereas for the
FM pattern the magnetism vanishes beyond a critical level of
doping. However, this feature for the FM spin pattern is of no
significance, because energetically it lies above both of the AFM
patterns at all levels of doping.

\subsection{{\bf Electric field gradient}}

Nuclear magnetic resonancce (NMR) is a local probe that is extremely
sensitive to certain details of the structure. Since the As $z$
position is a key determinant of many of the electronic properties of
the FeAs systems, the quadrupole frequency $\nu_{Q}$ from NMR
measurements can provide a direct measure to the Fe-As interaction.
Theoretically, $\nu_{Q}$ can be obtained by calculating the electric
field gradient (EFG).  The EFG is defined as the second partial
derivative of the electronic potential $v(\vec r)$ at the position of
the nucleus
\begin{eqnarray}
V_{ij}&=&\left(\partial_i\partial_j v(0) - \frac{1}{3}\Delta \delta_{ij} \right)
\Delta v(0).
\end{eqnarray}
This traceless and symmetric tensor of rank 2 is described in the
principal axis system by the main component $V_{zz}$ and the asymmetry
parameter $\eta = (V_{xx}-V_{yy})/V_{zz}$.  $V_{zz}$ is per definition
the component with the largest magnitude $|V_{zz}|\geq|V_{yy}| \geq
|V_{xx}|$ and is not necessarily parallel to the $z$-axis of the
crystal.  From these two parameters ($V_{zz}$ and $\eta$) and the
quadrupole moment for $^{75}$As $Q=(0.314\pm0.006)$~b \cite{pyykko}
the quadrupole frequency $\nu_Q$ for $^{75}$As (with a nuclear spin of
$I=3/2$) can be calculated \cite{abragam}
\begin{eqnarray}
\label{nuQ}
\nu_Q=\frac{eQV_{zz}}{2h}\sqrt{1+\frac{\eta^2}{3}}.
\end{eqnarray}

\begin{figure}[htb]
\begin{center}
\includegraphics[clip,width=8cm,angle=0]{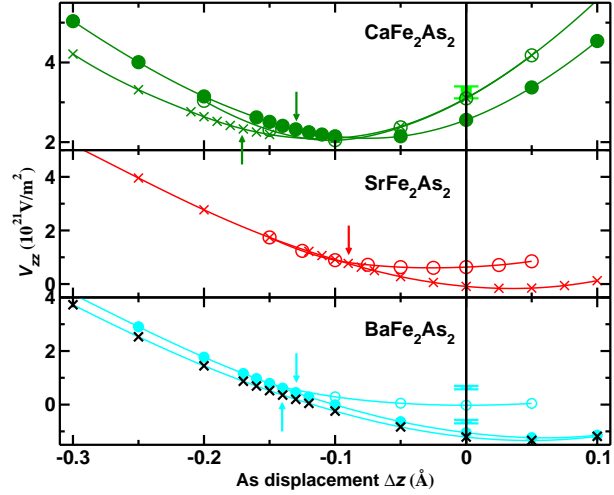}
\end{center}
\caption{\label{EFG1}Dependence of the EFG for As on the As $z$
  position. $\Delta z = z - z_{\mathrm{exp}}$. Different symbols show
  different calculations: cross = nonmagnetic with tetragonal
  symmetry, full circle = nonmagnetic with orthorhombic symmetry
  (almost identical to tetragonal symmetry for SrFe$_2$As$_2$ and
  therefore not shown) and empty circle = magnetic (SDW pattern) with
  orthorhombic symmetry. The total energy minimum is marked by an
  arrow for each nonmagnetic curve. The error bars show the
  experimental results for the tetragonal phase (at 250~K for
  CaFe$_2$As$_2$ and at 200~K for BaFe$_2$As$_2$ (the two error bars
  indicate the experimentally unknown sign of the EFG)).}
\end{figure}

The experimental lattice parameters, including the As $z$ position for
the calculation of the EFG for As in CaFe$_2$As$_2$, SrFe$_2$As$_2$
and BaFe$_2$As$_2$ were obtained from Ref.  \cite{Kreyssig08a},
\cite{Sr122GiPa}, \cite{Rotter08a} respectively. For the parent
compounds we investigated the influence of the As $z$ position, the
structural phase transition, the magnetism and the pressure on the
EFG. We also investigated the effects of doping on the EFG.

First we focus on the As $z$ dependence of the EFG.  Just as the
computed Fe magnetic moment, whose value show a strong dependence on
the As $z$ position (see Sec.~\ref{Aszpos}), the EFG is also found to
display a strong As $z$ dependence.  The EFG increases strongly for
all three compounds, as the Fe-As distance decreases, see
Fig.~\ref{EFG1}. In case of CaFe$_2$As$_2$, there is a minimum in the
EFG for a displacement of roughly $\Delta z=-0.1$~{\AA} from the
experimental position, while for larger Fe-As distances the EFG
increases again. The same trend is observed for the other two
compounds, middle and lower panels in Fig.~\ref{EFG1}. They exhibit
the minimum in the EFG at about $\Delta z=+0.05$~{\AA}.  For all the
three parent compounds, the Fe-As distance at which the minimum in
total energy occurs is smaller than that corresponds to the EFG
minimum (see arrows in Fig.~\ref{EFG1}).  For CaFe$_{2}$As$_{2}$ we
observe a good agreement between the calculated EFG at the
experimental As $z$ position (for 250~K) and the measured EFG at 250~K
\cite{Baek} (the light green symbol in the upper panel of
Fig.~\ref{EFG1}).  In case of BaFe$_2$As$_2$ the magnitude of the
measured EFG at 200~K is roughly $0.7\times 10^{21}$~V/m$^2$
\cite{Kitagawa}, while the sign is unknown since it cannot be
extracted from NQR measurements (In Fig.~\ref{EFG1} the experimental
EFG values with both signs are shown).  The calculated $V_{zz}$ for
the experimental As $z$ position is $-1.1\times 10^{21}$~V/m$^2$. If
the experimental EFG is negative, reasonable agreement between
experiment and calculation is obtained.  In a preliminary measurement
for SrFe$_{2}$As$_{2}$ the quadrupole frequency $\nu_Q$ was determined
to be positive and less than 2~MHz \cite{peter} and this is also
consistent with the calculated EFG of 0.8~MHz at the experimental As
$z$ position.  For members of the $1111$ family; LaFeAsO and NdFeAsO:
the calculated EFG for the optimized As $z$ position agreed better
with the experimental EFG \cite{Lapaper,Ndpaper}.  Our results for
three representative members of the $122$ family as shown above follow
a different trend: the calculated EFG using the experimental As $z$
position agree better with the measured EFG values.

To study the influence of the orthorhombic distortion, but without the
influence of magnetism we perform non-magnetic calculations both in
tetragonal and orthorhombic symmetry. The orthorhombic splitting of
the axes in the $(a,b)$ plane has a rather small influence on the
EFG. The EFG is larger for the orthorhombic symmetry for small Fe-As
distances, i.e. $\Delta z<-0.1$~{\AA}.  In case of SrFe$_2$As$_2$, the
effect of the orthorhombic splitting is so small, that the
orthorhombic EFG curve in Fig.~\ref{EFG1} is not shown (see middle
panel of Fig.~\ref{EFG1}). In case of BaFe$_2$As$_2$, we observe
similar behaviour as for CaFe$_2$As$_2$. The tetragonal and
orthorhombic EFG curves cross close to the EFG minimum and the EFG is
larger for the orthorhombic symmetry for smaller Fe-As distances,
i.e. $\Delta z<+0.1$~{\AA}.  For all three compounds we find that
$V_{zz}$ is parallel to the crystallographic $z$-axis for the
non-magnetic calculations in both the tetragonal and orthorhombic
symmetry.

\begin{figure}[b]
  \begin{center}
    \begin{minipage}[t]{0.48\linewidth}
      \epsfig{file=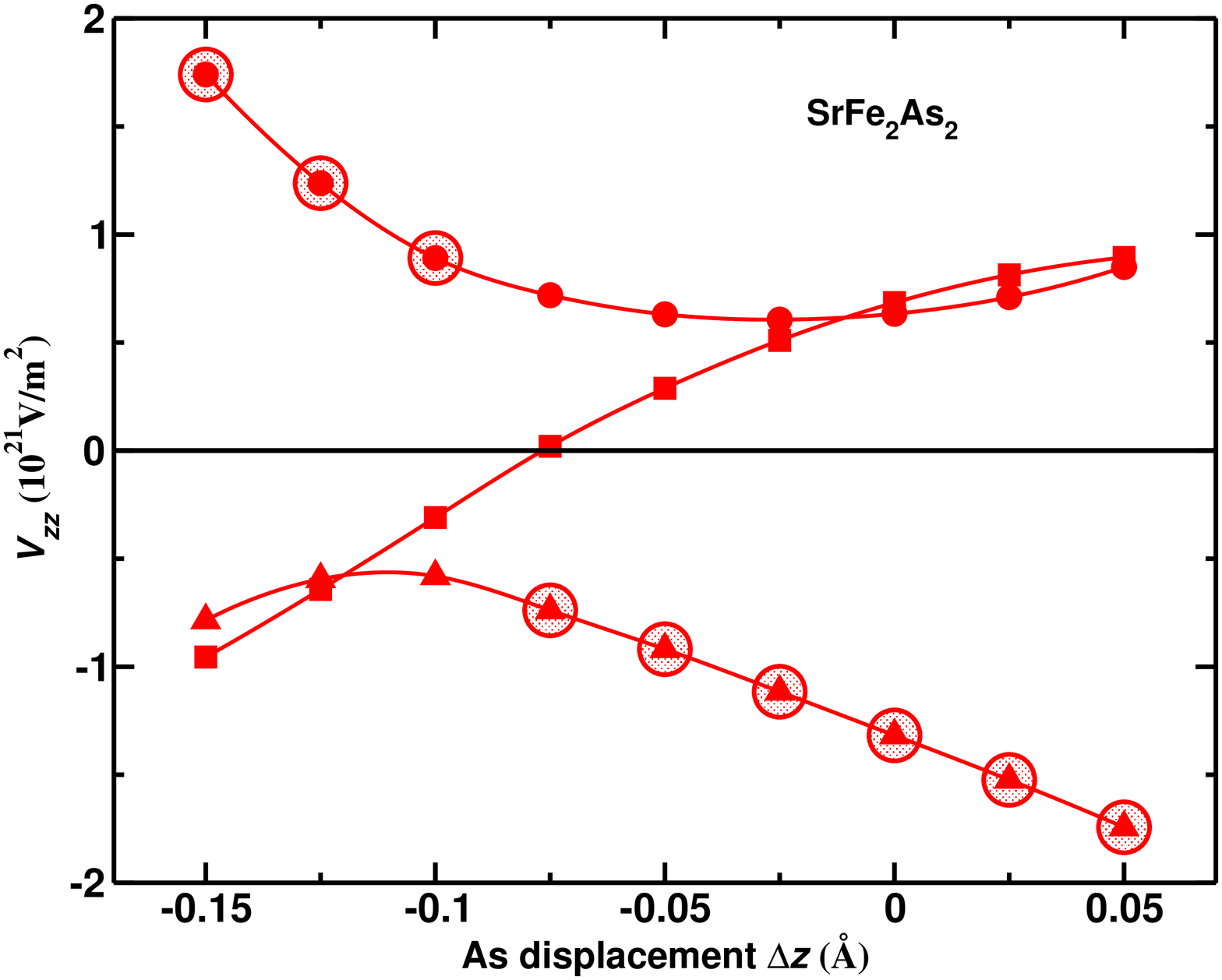, clip=, angle=-90, width=\linewidth}
    \end{minipage}
    \begin{minipage}[t]{0.48\linewidth}
      \epsfig{file=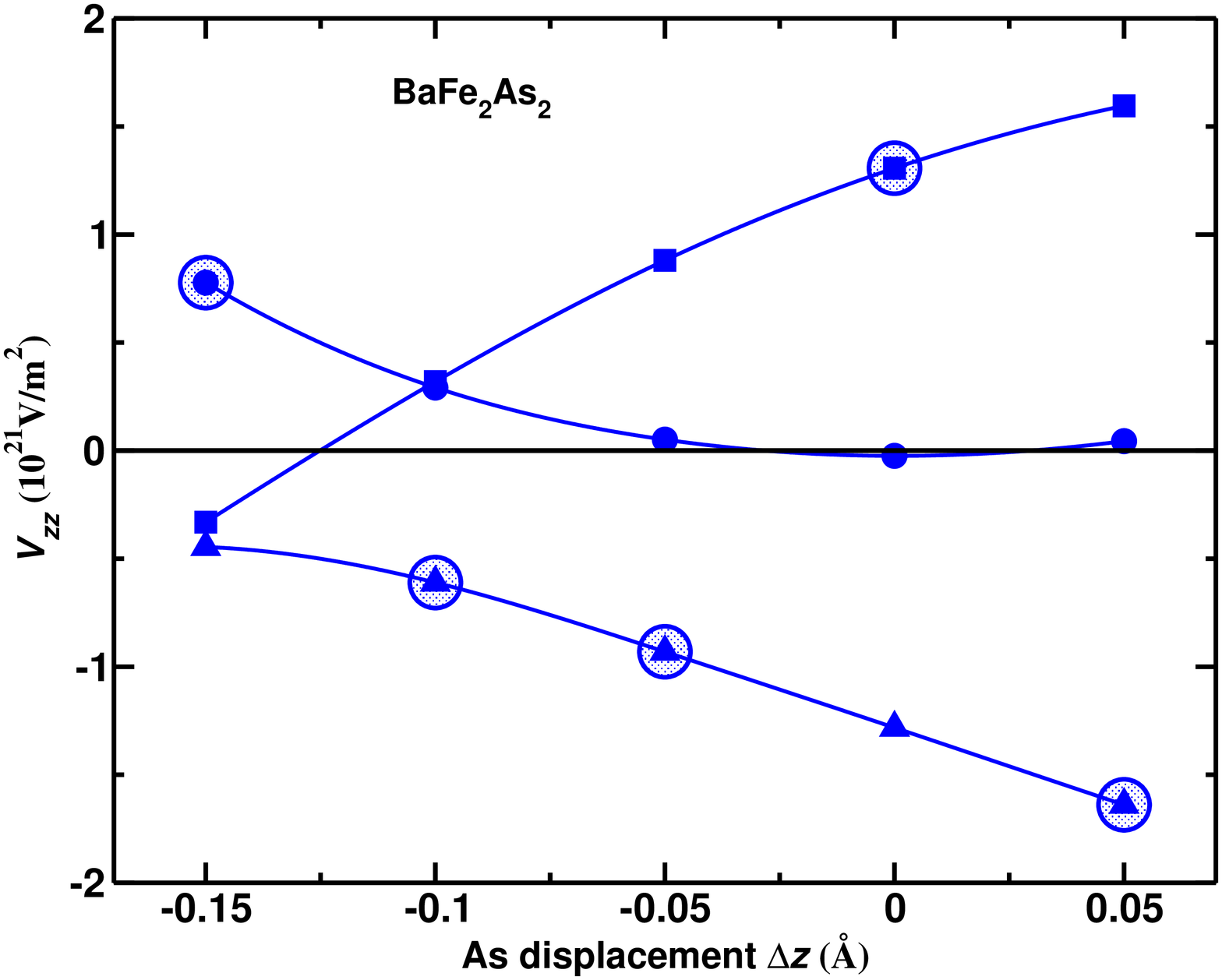, clip=, angle=-90, width=\linewidth}
    \end{minipage}
\caption{\label{EFG2}{\bf Left:} The three components of the EFG
  tensor for SrFe$_2$As$_2$ in the orthorhombic SDW phase as a
  function of the As $z$ position. $\Delta z = z -
  z_{\mathrm{exp}}$. The component of the EFG, parallel to the
  crystallographic $x$-axis is shown by triangles up, the component
  parallel to the crystallographic $y$-axis by squares and the one
  parallel to the crystallographic $z$-axis by solid
  circles. $V_{zz}$, the largest one of these three, is marked by a
  large shaded circle for each As $z$ position.  {\bf Right:} The
  three components of the EFG tensor for BaFe$_2$As$_2$ in the
  orthorhombic SDW phase as a function of the As $z$ position. The
  rest of the notation is the same as the left panel. }
  \end{center}
\end{figure}

\begin{table}[t]
\begin{center}
\caption{$V_{zz}$ in $10^{21}$~V/m$^2$ for the nonmagnetic and
  different magnetic orders, all in orthorhombic phase.}\label{EFGtable}
\begin{tabular}{lcccc}
 compound &  NM &  FM & NN--AFM  & SDW
\\
\hline
CaFe$_2$As$_2$ & 2.6 & 2.4 & 2.7 & 3.1
\\
SrFe$_2$As$_2$ & 0.2 & 0.3 & 0.2 &-1.3
\\
BaFe$_2$As$_2$ & -1.1 & -1.0 & -1.3 &+1.3
\\
\hline
\end{tabular}
\end{center}
\end{table}

Investigation of the influence of the magnetism on the EFG in the
orthorhombic symmetry show that, FM or NN--AFM ordering of the Fe
atoms does not change the EFG much, however, the SDW order has a huge
influence on the EFG, \textit{cf.} Table~\ref{EFGtable}. For all the
three systems, as the Fe-As distance is decreased, the magnetic moment
is reduced and finally tends to zero. At this displacement value, the
SDW EFG curves smoothly join the non-magnetic orthorhombic EFG curves
as one would expect.  In Fig.~\ref{EFG1} the component of the EFG,
that is parallel to the $z$-axis of the crystal is shown. As mentioned
before, $V_{zz}$ is found to be parallel to the crystallographic
$z$-axis in all non-magnetic calculations. For the magnetic SDW phase
the same behaviour is observed for CaFe$_2$As$_2$, but not for
SrFe$_2$As$_2$ and BaFe$_2$As$_2$.  For the latter two compounds,
$V_{zz}$ changes the axis, \textit{i.e.}, the axis along which EFG is
the largest changes as Fe-As distance is varied. For
BaFe$_{2}$As$_{2}$, such a behaviour was also observed experimentally
\cite{Kitagawa} when going from the high-temperature nonmagnetic
tetragonal phase to the low temperature SDW phase.  Unfortunately for
CaFe$_2$As$_2$, only the quadrupole frequency parallel, and not
perpendicular to the crystallographic $z$-axis is provided in
Ref.~\cite{Baek}. The three diagonal components of the EFG tensor
$V_{ii}$, which are parallel to the $x$, $y$ and $z$-axis of the
crystal, vary continuously as a function of the As $z$ position, as
can be seen in Fig.~\ref{EFG2}. In case of SrFe$_2$As$_2$, $V_{zz}$ is
parallel to the $x$-axis for a displacement of As between
$+0.05$~{\AA} and $-0.075$~{\AA} (which includes the experimental As
$z$ position) and parallel to the $z$-axis for a displacement between
$-0.1$~{\AA} and $-0.15$~{\AA}.
For BaFe$_2$As$_2$ according to its definition as the largest
component $V_{zz}$ fluctuates between all the three different axes
(Fig.~\ref{EFG2}) In particular, at the experimental As $z$ position
$V_{zz}$ is parallel to the $y$-axis.  We also observe that the
component parallel to the $x$-axis is very similar for both Sr and Ba
122 compounds.  The components parallel to the $y$- and $z$-axis show
the same variation with As $z$ position, only the values for the two
compounds are shifted by an almost constant amount.
Fig.~3 in Ref.~\cite{Baek} shows the temperature dependence of the
quadrupole frequency $\nu_Q$ for CaFe$_2$As$_2$: $\nu_Q$ increases
drastically from 300~K to 170~K. At 170~K there is a large jump in the
frequency due to the orthorhombic SDW phase transition. Between 170~K
and 20~K $\nu_Q$ is rather constant.  The calculated EFGs correspond
to lattice parameters at 250~K and 50~K. For these two temperatures
the quadrupole frequency (parallel to the crystallographic $z$-axis)
is almost identical. This is in agreement with our result for the
experimental As $z$ position as seen in the upper panel of
Fig.~\ref{EFG1}.

\begin{figure}[htb]
\begin{center}
\includegraphics[clip,width=8cm,angle=-90]{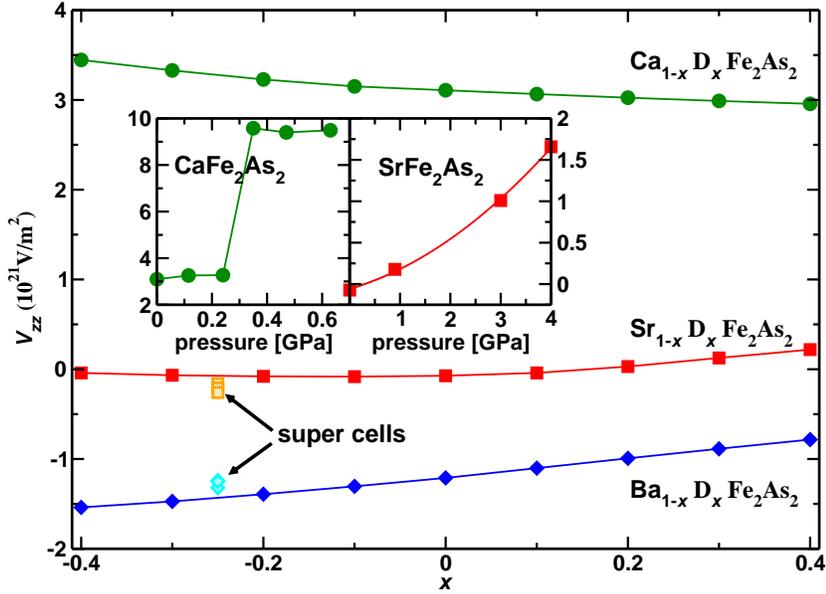}
\end{center}
\caption{\label{EFG3} EFGs calculated for doped CaFe$_2$As$_2$ (green
  circles), SrFe$_2$As$_2$ (red squares) and BaFe$_2$As$_2$ (blue
  diamonds) using VCA. Results obtained from a four-fold super cell
  for {\bf $x$} = -0.25 for SrFe$_2$As$_2$ (shaded orange squares) and
  BaFe$_2$As$_2$ (shaded blue diamonds) in the nonmagnetic tetragonal
  phase are also shown. {\bf Inset:} Dependence of the EFG on pressure
  for CaFe$_2$As$_2$ (left) and SrFe$_2$As$_2$ (right). The latter is
  in the (nonmagnetic) tetragonal phase. }
\end{figure}

We also investigated the influence of pressure on the EFG.  $V_{zz}$
for different pressures was calculated for CaFe$_2$As$_2$
\cite{Kitagawa} and SrFe$_2$As$_2$ \cite{Kumar08a} using the
experimental structural parameters reported as a function of pressure.
Our result is shown in the inset of Fig.~\ref{EFG3}. In case of
CaFe$_2$As$_2$, the EFG increases when the applied pressure is
increased from 0~GPa to 0.24~GPa. For these pressures the structure is
in the (orthorhombic) SDW phase. The next experimental pressure point
is larger than the critical pressure of 0.3~GPa (Sec.~\ref{pressure}),
where the $c/a$ collapse takes place. The structure changes into the
nonmagnetic tetragonal phase and the calculated EFG increases
drastically from roughly 3 to 10$\times 10^{21}$~V/m$^2$.
Experimentally, the applied pressure for SrFe$_2$As$_2$ was much
higher (up to 4~GPa) than for CaFe$_2$As$_2$, but no indications of a
collapsed phase was found till now. Contrary to the jump in the
calculated EFG at 0.3 GPa for CaFe$_{2}$As$_{2}$, EFG for
SrFe$_{2}$As$_{2}$ increases monotonously without any kinks upto 4
GPa.  It is worthwhile to measure the EFG for these systems to get a
more clear picture.

Finally, the EFGs of the $A$ site doped compounds were calculated with
VCA. The validity of the VCA was checked by super cell calculations
for SrFe$_2$As$_2$ and BaFe$_2$As$_2$. Due to the super cell
construction, there are three different Wyckoff positions for As and
hence three different EFGs, which lie reasonably close to the VCA EFG
curve.  In the VCA calculation we keep the structural parameters fixed
for the different levels of doping. In Fig.~\ref{EFG3} EFGs calculated
in this manner are shown for CaFe$_2$As$_2$, SrFe$_2$As$_2$ and
BaFe$_2$As$_2$.  In case of CaFe$_2$As$_2$, the EFG increases when
electrons are taken out and decreases when electrons are added to the
system. This implies that the As electron density gets more isotropic,
when the system is electron doped.  For BaFe$_{2}$As$_{2}$, the trend
is the same as in CaFe$_{2}$As$_{2}$ whereas for SrFe$_2$As$_2$ the
situation is slightly different: hole doping does not change the EFG
much, while electron doping increases the EFG. Note however, that the
calculated EFG for SrFe$_{2}$As$_{2}$ are quite small.

We conclude that the EFG in the 122 and in the 1111 systems behave
similar \cite{Lapaper,Ndpaper}: the effect of electron doping on the
EFG is much smaller than the influence of the As $z$ position and
pressure (compare Fig.~\ref{EFG1} and Fig.~\ref{EFG3}).  This finding
emphasizes again the crucial importance of a correct description of
the Fe-As interaction (that is mostly responsible for the density
around As) for the physical properties of the iron pnictides.

\section{{\bf Results - Experiment}}

\subsection{{\bf Substitutions of Fe by other 3$d$-metals}}

As already mentioned, at ambient pressure, no bulk superconductivity
has been observed in stoichiometric $RE$OFeAs and $A$Fe$_2$As$_2$ ($A$
= Ca, Sr, Ba, Eu) compositions (except for one report on
SrFe$_2$As$_2$ crystals \cite{Saha08a}). Instead, these parent
compounds display the SDW transition at typically 100--200\,K. A
modification of the intralayer $d^\mathrm{Fe-Fe}_a$ and interlayer
$d^\mathrm{Fe-Fe}_c$ distances and thus of the electronic states at
$E_F$ can be achieved by different means: (i) application of
hydrostatic (or uniaxial) pressure, (ii) isovalent substitution of a
constituent atom by a smaller/larger ion in order to apply chemical
pressure, (iii) hole doping or (iv) electron doping by non-isovalent
substitution of any of the constituent atoms. The latter two methods
usually also excert chemical pressure. Most excitingly, all these
methods have been proven to be successful in generating bulk
superconductivity in iron arsenide systems.

Application of pressure has widely been used to explore the phase
diagrams of the superconducting chemical systems
\cite{Torikachvili08a,Park08a,Alireza08a}. The method usually introduces
no crystallographic disorder in the structural building units. In
contrast when
substituting a chemical constituent (methods ii--iv) always a certain
degree of structural disorder is introduced. First, only substitutions
on sites \textit{in-between the Fe-As layers} were attempted. Based on
the experience gained from extensive work on cuprates, such an indirect
doping of the Fe-As layers is expected to introduce only minor
structural disorder. In a localized (cuprate-like) as well as in an
itinerant model of the arsenides, this type of doping amounts to a
simple charge doping. Such experiments are therefore not suitable for
discriminating between both models.

In contrast, a substitution of an atom species \textit{within the
  Fe-As layer} can yield more information on the underlying
physics. In an itinerant model the substitution of a small amount of
Fe by another $d$ element ($TM$) is expected to be similar to indirect
doping since only the total count of electrons is relevant, i.e.\ a
rigid-band picture should work in first approximation. In a picture
with localized $d$ electrons, on the other hand, doping on the Fe site
should directly affect the correlations in the Fe-As layers. A
behaviour drastically different from indirect doping should evolve. In
cuprates the substitution of a few percent Ni or Zn on the Cu site
leads to a strong reduction of $T_c$.

Therefore, several groups recently investigated the properties of
solid solutions of the type $RE$O(Fe$_{1-x}$Co$_x$)As or
$A$Fe$_{2-x}$Co$_x$As$_2$. Sefat \textit{et al.}\,\cite{Sefat08aetal}
and Wang \textit{et al.}\,\cite{CWang08aetal} first reported
superconductivity in cobalt doped LaOFeAs with a maximum $T_c \approx
10$\,K. Our group concentrated on the system SrFe$_{2-x}TM_x$As$_2$:
while the pure Fe compound undergoes a lattice distortion and SDW
ordering at $T_0$ = 205\,K \cite{Krellner08a}, Co substitution leads
to a rapid decrease of $T_0$, followed by the onset of bulk
superconductivity in the concentration range $0.2 \leq x \leq 0.4$
\cite{LeitheJasper08b}. The maximum $T_c$ of $\approx 20$\,K is
achieved for $x \approx 0.20$. This was in fact also the first
observation of \textit{electron-doping} induced superconductivity in
$A$Fe$_2$As$_2$ compounds. Co substitution also generated bulk
superconductivity with maximum $T_c \approx 22$\,K in
BaFe$_{2-x}$Co$_x$As$_2$ \cite{Sefat08b}, however, the optimal doping
seems to be lower than in the Sr system
\cite{Tanatar08a}. Substitution of the following $TM$, nickel,
introducing twice as many electrons per atom into the Fe-As layer,
also generates bulk superconductivity, albeit with lower $T_c$ than Co
substitution in the Sr compound \cite{LeitheJasper08b}. However, for
the corresponding Ba compound $T_c$ up to 21\,K is reported for
BaFe$_{1.90}$Ni$_{0.10}$As$_2$ \cite{LJLi08a}. Only very recently
another internal substitution, namely of As by P, was reported for
EuFe$_2$As$_2$ \cite{ZRen08a} and LaOFeAs \cite{CWang08c}. Also by
this means the SDW transition can be influenced and superconductivity
can be induced.

\begin{figure}[htb]
\begin{center}
\includegraphics[height=0.8\textwidth,angle=90]{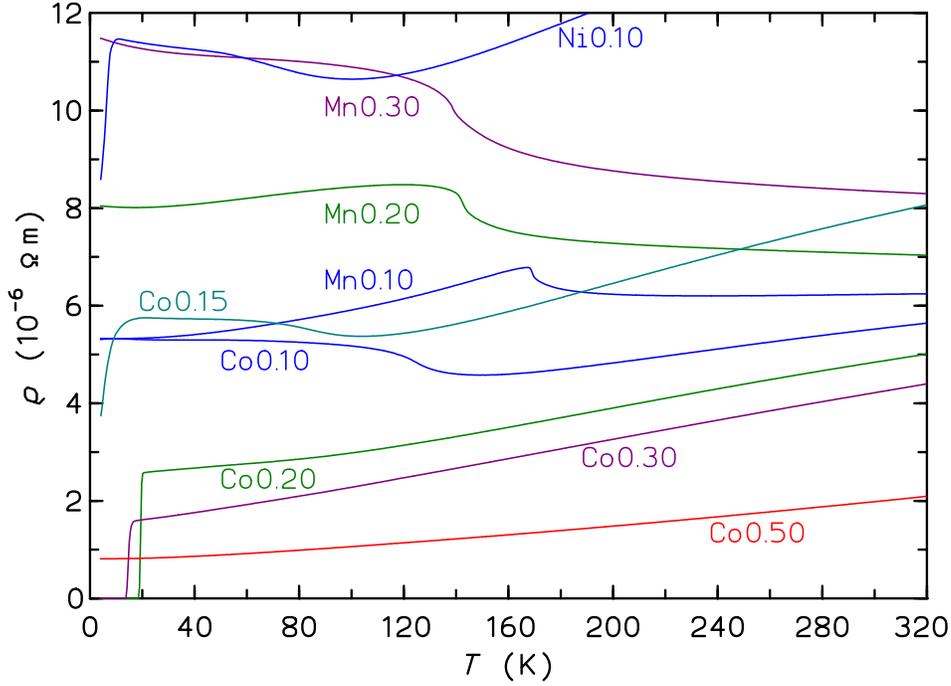}
\caption{
Electrical resistivity of polycrystalline
SrFe$_{2-x}TM_x$As$_2$ ($TM$ = Mn, Co, Ni) samples
\label{figrho}}
\end{center}
\end{figure}

In Table \ref{tableTc} we present the lattice parameters and the SDW
and superconducting transition temperatures ($T_0$, $T_c$) of several
SrFe$_{2-x}TM_x$As$_2$ solid solutions. Values for $T_0$ can be most
easily obtained from the corresponding anomaly in resistivity data
(see Fig.\ \ref{figrho}). Besides Co and Ni substitutions in the Sr122
and Ba122 systems, no further $d$ element substitutions have been
reported yet. As demonstrated recently \cite{LeitheJasper08b},
substitution of Fe by Co suppresses rapidly $T_0$ (see
Fig.\ \ref{figrho}). Bulk superconductivity, as proven by specific
heat, magnetic shielding, and resistivity data, appears when $T_0 = 0$
or $T_0 < T_c$, which is reached for $x > 0.20$
\cite{LeitheJasper08b}. Only about half of the substituting element
($x \approx 0.10$) is necessary to induce bulk superconductivity when
using nickel \cite{LJLi08a}. Both elements then introduce 0.2 excess
electrons into the FeAs layers. While the $a$ lattice parameter does
not change significantly with Co substitution the $c$ lattice
parameter decreases continuously in SrFe$_{2-x}$Co$_x$As$_2$
\cite{LeitheJasper08b} and with Ni content in BaFe$_{2-x}$Ni$_x$As$_2$
\cite{LJLi08a}. Chemical homogeneity of the Co or Ni distribution is
still an issue in current samples. One of the most important
questions, the co-existence or mutual exclusion of SDW state and bulk
superconductivity, bas been discussed heavily for
$A$Fe$_2$As$_2$-based alloy series
\cite{Rotter08c,JHChu08a,XFWang08a}. The answer is currently open and
can be only given for really homogeneous samples.

Direct, in-plane \textit{hole} doping might also induce
superconductivity. Our new investigations show, that a substitution of
Fe by Mn is possible and that it leads to a continuous increase of
both the $a$ and $c$ lattice parameters with Mn content. Under these
conditions a hole doping does not generate superconductivity. In
contrast, in the indirectly-doped Sr$_{1-x}$K$_x$Fe$_2$As$_2$
\cite{Sasmal08a,GFChen08aetal} and Ba$_{1-x}$K$_x$Fe$_2$As$_2$
\cite{Rotter08b} compounds the $a$ lattice parameter decreases with
$x$ while $c$ increases, keeping the unit cell volume almost
constant. Also, the SDW transition temperature $T_0$ is suppressed
with increasing Mn content in a different way (see Fig.\ \ref{figrho})
than for the Co and Ni substitutions where $T_0$ is suppressed for $x
\approx 0.20$ and $x \approx 0.10$, respectively. Indeed, after an
initial decrease of $T_0$ to $\approx$140\,K for $x = 0.20$ the
transition temperature seems not to decrease further with $x$. Thus,
the doping with holes or electrons is not the only important factor
for the appearance of superconductivity but a corresponding tuning of
the distances $d^\mathrm{Fe-Fe}_a$ and $d^\mathrm{Fe-Fe}_c$ has to be
accomplished also. At present, the microscopic origin of the
differences upon Mn doping compared to Co or Ni doping is unclear.
Therefore, further investigations for different substitutions are
currently underway.

\begin{table}[htb]
\begin{center}
\caption{Lattice parameters $a,c$ of some SrFe$_{2-x}TM_x$As$_2$
  (nominal compositions) phase and superconducting transition
  temperature $T_c^\mathrm{mag}$ from magnetization measurements
  (onset, $T_\mathrm{min} = $1.8\,K).
\label{tableTc}}
\begin{tabular}{lcccccl}
$TM$& $x$  & $a$        & $c$        &$T_c^\mathrm{mag}$& $T_0$     & Ref.                   \\
    &      & (\AA)      & (\AA)      & (K)              & (K)       &                        \\ \hline
--  & 0.00 & 3.924(3)   & 12.38(1)   & --               & 205       & \cite{Krellner08a}     \\ \hline
Co  & 0.10 & 3.9291(1)  & 12.3321(7) & --               & 130       & \cite{LeitheJasper08b} \\
    & 0.15 & 3.9272(1)  & 12.3123(5) & --               &  90       & \cite{LeitheJasper08b} \\
    & 0.20 & 3.9278(2)  & 12.3026(2) & 19.2             & \textless~30     & \cite{LeitheJasper08b} \\
    & 0.25 & 3.9296(2)  & 12.2925(9) & 18.1             & --        & \cite{LeitheJasper08b} \\
    & 0.30 & 3.9291(2)  & 12.2704(8) & 13.2             & --        & \cite{LeitheJasper08b} \\
    & 0.40 & 3.9293(1)  & 12.2711(7) & 12.9             & --        & \cite{LeitheJasper08b} \\
    & 0.50 & 3.9287(2)  & 12.2187(9) & --               & --        & \cite{LeitheJasper08b} \\
    & 2.00 & 3.9618(1)  & 11.6378(6) & --               & --        & \cite{LeitheJasper08b} \\ \hline
Ni  & 0.10 & 3.9299(1)  & 12.3238(6) & $\approx 8$      & \textless~85      & \cite{LeitheJasper08b} \\ \hline
Mn  & 0.10 & 3.9319(2)  & 12.4161(7) & --               & 165       & this work              \\
    & 0.20 & 3.9384(3)  & 12.4615(23)& --               & 130       & this work              \\
    & 0.30 & 3.9441(2)  & 12.4832(7) & --               & 130       & this work              \\
\end{tabular}
\end{center}
\end{table}
\noindent

Magnetic susceptibility and specific heat data for polycrystalline
samples SrFe$_{2-x}$Co$_x$As$_2$ have already been presented
\cite{LeitheJasper08b}. Here, instead we report newer data obtained on
two crystals with Co contents $x > 0.2$ grown by a flux method (see
section \ref{experimental}).

Both crystals show strong diamagnetic signals in measurements after
zero field cooling (ZFC). The onset temperatures $T_c^\mathrm{mag}$
are 17.8 K and 15.3\,K, respectively. While the transition for crystal
X1 is much wider than that of crystal X2 the $T_c$ of the latter is
somewhat lower, indicating a slightly larger Co-content in accordance
with the EPMA investigations. The shielding signals (ZFC) corresponds
to the whole sample volume, however the Meissner effect (FC) is very
small. This is typical for Co-substituted $A$Fe$_2$As$_2$ materials
and probably due to strong flux line pinning. The random (and somewhat
inhomogeneous) substitution of Fe by Co within the superconducting
layers seems to introduce effective pinning centers. This is a
remarkable difference to superconducting compositions with
substitutions outside the Fe-As layers.

As already demonstrated \cite{LeitheJasper08b}, the SDW ordering and
the connected lattice distortion at $T_0$ = 205\,K in SrFe$_2$As$_2$
\cite{Krellner08a,Tegel08a} is strongly suppressed by Co substitution,
similar as for K substitution (indirect hole doping)
\cite{GFChen08aetal}. For the two crystals no corresponding anomaly in
$\rho(T)$ is observed.

\begin{figure}[htb]
\begin{center}
\includegraphics[height=0.8\textwidth,angle=90]{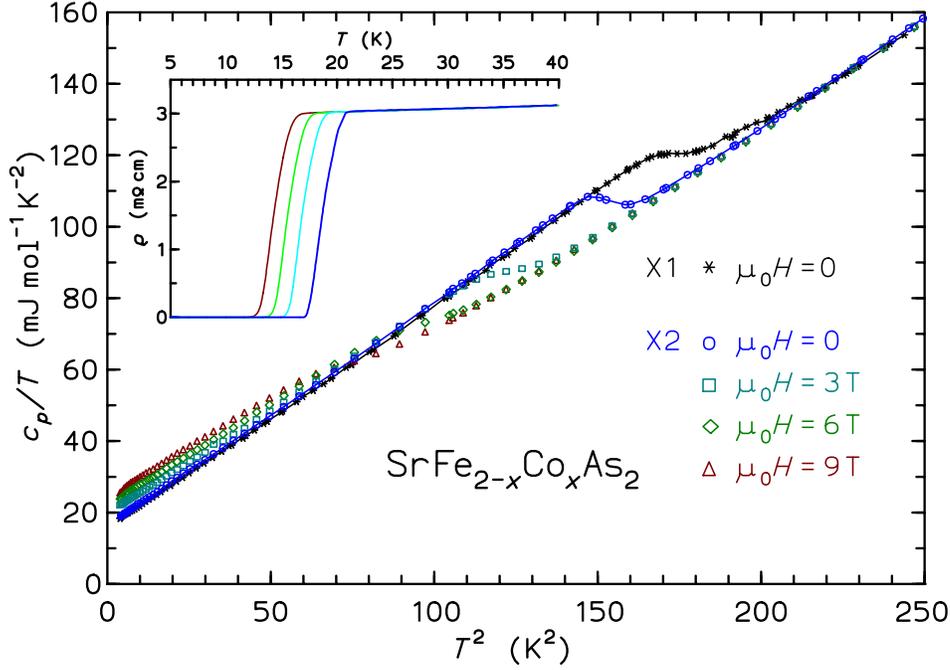}
\caption{Molar isobaric specific heat $c_p/T^{2}$ of
SrFe$_{2-x}$Co$_x$As$_2$ crystal samples for different magnetic fields.
For crystal X1 only data for $\mu_0H = 0$ are given. Zero-field data
points are connected by a line. The inset shows the resistivity of
crystal X2 around $T_c$ for $\mu_0H$ = 0.01, 3, 6, and 9\,T.
\label{figcp}}
\end{center}
\end{figure}

The specific heat $c_p(T)$ for the two crystals are shown in
Fig.\ \ref{figcp} in a $c_p(T)/T^{2}$ versus $T$ representation. It
can be clearly seen that crystal X1 has a very broad transition (with
a ``foot'' at the high-temperature side) while crystal X2 displays a
rounded but single-step anomaly. The specific heat jumps $\Delta
c_p/T_c$ and the transition temperatures $T_c^\mathrm{cal}$ can be
evaluated by a fit including a phonon background (harmonic lattice
approximation) and an electronic term $c_{es}(T)$ according to the BCS
theory ($\Delta c_p/T_c = 1.43\gamma$) or the phenomenological
two-liquid model ($\Delta c_p/T_c = 2\gamma$, a model for stronger
e-ph coupling). The inclusion of a residual linear term $\gamma'T$ was
found to be absolutely necessary for a good fit.
$$
c_p(T) = \gamma' \: T + \beta T^{3} + \delta T^{5} + c_{es}(T)
$$
The jump at $T_c$ is ``broadened'' in order to simulate the
rounding of the transition steps due to chemical inhomogeneities. For
the more homogeneous sample X2 no difference in the least-squares
deviation is observed between the BCS and two-fluid model. We find for
crystal X2 $\Delta c_p/T_c^\mathrm{cal} \approx$12.0 mJ mol$^{-1}$
K$^{-2}$ at $T_c^\mathrm{cal}$ = 12.34\,K and also for crystal X1
similar values ($\approx$12.5 mJ mol$^{-1}$ K$^{-2}$; $T_c$ = 13.22\,K).
Similar values of $\Delta c_p/T_c^\mathrm{cal}$ were observed in our
previous study \cite{LeitheJasper08b} of polycrystalline samples with
Co-contents $x = 0.20$ and $x = 0.30$.

The existence of a linear specific heat term well below $T_c$ is found
for several $A$Fe$_2$As$_2$-based alloys
\cite{LeitheJasper08b,GMu08a}.  For $H = 0$ we observe for both
crystals values of $\gamma'$ around 20 mJ mol$^{-1}$
K$^{-2}$. $\gamma'$ generally increases with field.  Whether the
residual $\gamma'$ is due to defects as in the case of early cuprate
superconductor samples (see e.g. Ref.\,\cite{TrisconeJunod96}) or
whether it is an intrinsic contribution has to be clarified by further
experiments. An intrinsic reason could be some ungapped parts of the
Fermi surface \cite{Drechsler03aetal}. For the specific heat jump
$\Delta c_p/T_c$ in BaFe$_{2-x}$Co$_x$As$_2$ also relatively small
values are reported ($\approx$ 25 mJ mol$^{-1}$ K$^{-2}$
\cite{Tanatar08a}) while for Ba$_{0.6}$K$_{0.4}$Fe$_2$As$_2$ ($\Delta
c_{p}/T_c \approx$ 100 mJ mol$^{-1}$ K$^{-2}$ \cite{GMu08a,Rotter08c})
much larger jumps are observed. This may indicate that the
superconducting Fermi surface portions in the Co-substituted compounds
(in-plane doping) are strongly different from those in indirectly
doped superconductors.  Recent photoemission (ARPES) investigations
indeed point out severe differences between (non-superconducting)
BaFe$_{1.7}$Co$_{0.3}$As$_2$ and (superconducting)
Ba$_{0.6}$K$_{0.4}$Fe$_2$As$_2$ \cite{Sekiba08a}.  In conclusion, our
doping experiments on the Fe site strongly favor an intinerant picture
over a localized scenario.  Further thermodynamic and Fermi surface
studies are required to resolve this issue.

\section{\bf Summary}

In this paper, we presented a joint theoretical and experimental study
of the systems $A$Fe$_{2}$As$_{2}$ ($A$ = Ca, Sr, Ba, Eu) and
SrFe$_{2-x}$$TM_{x}$As$_{2}$ ($TM$ = Mn, Co, Ni) to investigate the
relation of crystal structure and charge doping to magnetism and
superconductivity in these compounds.  Based on {\it ab-initio}
electronic structure calculations we focused on the relationship
between the crystal and electronic structure, charge doping and
magnetism for the 122 family since their electronic structure and
physical properties are quite similar  to the other superconducting
iron pnictide families (1111, 1111', 111, 11).

Although problems with an accurate description of the Fe-As
interaction persist in present-day density functional calculations,
this approach provides deep inside into many questions.  We
demonstrated that tetragonal to orthorhombic transition in the 122
compounds is intrinsically linked to the SDW formation in agreement
with experimental observations. We find an anisotropic,
pressure-induced volume collapse for $A$Fe$_{2}$As$_{2}$ ($A$ = Ca,
Sr, Ba, Eu) that goes along with the suppression of the SDW magnetic
order. For $A$ = Ca our calculations are in excellent agreement with
the experimental observations \cite{Kreyssig08a}. An experimental
verification for the other compounds ($A$ = Ca, Sr, Ba, Eu) would be
desirable. With respect to the doping dependence in
SrFe$_{2-x}$$TM_{x}$As$_{2}$, we find the correct trends compared to
the experimental results. A more quantitative comparison will require
the explicit treatment of the influence of the substitutional disorder
on electronic structure and magnetism. This task is left for an
extended future study.

As demonstrated also for the EFG, many properties of these compounds
are sensitive to the As $z$ position. Since the magnetism and
therefore the superconductivity crucially depend on this structural
feature and the related accurate description of the Fe-As interaction,
the improvement of the calculations in this respect may offer the key
to the understanding of superconductivity in the whole family. In
order to improve the present-day density functional calculation for
this class of materials, the first step requires a deeper
understanding of where, how and why these DFT calculation
fail. Besides further calculational effort, a broader experimental
basis, especially high pressure studies will be necessary to approach
this complex issue.

Experimentally, we investigate the substitution of Fe in
SrFe$_{2-x}TM_{x}$As$_{2}$ by other 3$d$ transition metals, $TM$ = Mn,
Co, Ni.  In contrast to a partial substitution of Fe by Co or Ni
(electron doping) a corresponding Mn partial substitution does not
lead to the suppression of the antiferromagnetic order or the
appearance of superconductivity.  

The observed existence of a linear specific heat term in
SrFe$_{2-x}$Co$_{x}$As$_{2}$ well below $T_c$ is extremely important for
the understanding of the superconductivity in this compound and all
related 122 superconductors with doping on the Fe site. Therefore, a
careful investigation whether this feature is intrinsic or not is a
crucial question. In order to answer it, further studies on carefully
prepared and characterized high-quality samples are required. Since in
many experiments a considerable sample dependence is observed, this
issue is of large general importance for a future understanding of the
superconducting iron pnictide materials.

We thank T.\ Vogel, R.\ Koban, K.\ Kreutziger, Yu.\ Prots, and R.\
Gumeniuk for assistance.


\begin{thebibliography}{10}
\expandafter\ifx\csname url\endcsname\relax
  \def\url#1{{\tt #1}}\fi
\expandafter\ifx\csname urlprefix\endcsname\relax\def\urlprefix{URL }\fi
\providecommand{\eprint}[2][]{\url{#2}}

\bibitem{Kamihara08a}
Kamihara Y, Watanabe T, Hirano M and Hosono H 2008 {\em J.\ Am.\ Chem.\ Soc.\/}
  {\bf 130} 3296--7

\bibitem{Ren08a}
Ren Z~A, Lu W, Yang J, Yi W, Shen X~L, Li Z~C, Che G~C, Dong X~L, Sun L~L, Zhou
  F and Zhao Z~X 2008 {\em Chin. Phys. Lett.\/} {\bf 25} 2215

\bibitem{definition}
Our nomenclature of ``-{\bf $x$}" means hypovalent subsititution, while ``{\bf
  $x$}" means hypervalent substitution

\bibitem{Sasmal08a}
Sasmal K, Lv B, Lorenz B, Guloy A~M, Chen F, Xue Y and Chu C 2008 {\em Phys.\
  Rev.\ Lett.\/} {\bf 101} 107007

\bibitem{GFChen08a}
Chen G~F, Li Z, Li G, Hu W~Z, Dong J, Zhang X~D, Zheng P, Wang N~L and Luo J~L
  2008 {\em Chin. Phys. Lett.\/} {\bf 25} 3403

\bibitem{Rotter08b}
Rotter M, Tegel M and Johrendt D 2008 {\em Phys.\ Rev.\ Lett.\/} {\bf 101}
  107006

\bibitem{WangXC08a}
Wang X~C, Liu Q~Q, Lv Y~X, Gao W~B, Yang L~X, Yu R~C, Li F~Y and Jin C~Q 2008
  ArXiv:0806.4688v3

\bibitem{Mizuguchi08a}
Mizuguchi Y, Tomioka F, Tsuda S, Yamaguchi T and Takano Y 2008 {\em Appl.\
  Phys.\ Lett.\/} {\bf 93} 152505

\bibitem{Wu08b}
Wu G, Xie Y~L, Chen H, Zhong M, Liu R~H, Shi B~C, Li Q~J, Wang X~F, Wu T, Yan
  Y~J, Ying J~J and Chen X~H 2008  ArXiv:0811.0761v2

\bibitem{Ronning08a}
Ronning F, Klimczuk T, Bauer E~D, Volz H and Thompson J~D 2008 {\em J.\ Phys.:
  Condens.\ Matter\/} {\bf 20} 322201

\bibitem{Krellner08a}
Krellner C, {Caroca-Canales} N, Jesche A, Rosner H, Ormeci A and Geibel C 2008
  {\em Phys.\ Rev.\ B\/} {\bf 78} R100504

\bibitem{Rotter08a}
Rotter M, Tegel M, Schellenberg I, Hermes W, P{\"o}ttgen R and Johrendt D 2008
  {\em Phys.\ Rev.\ B\/} {\bf 78} R020503

\bibitem{Jeevan08a}
Jeevan H~S, Hossain Z, Kasinathan D, Rosner H, Geibel C and Gegenwart P 2008
  {\em Phys.\ Rev.\ B\/} {\bf 78} 052501

\bibitem{Jeevan08b}
Jeevan H~S, Hossain Z, Kasinathan D, Rosner H, Geibel C and Gegenwart P 2008
  {\em Phys.\ Rev.\ B\/} {\bf 78} 092406

\bibitem{LeitheJasper08b}
{Leithe-Jasper} A, Schnelle W, Geibel C and Rosner H 2008 {\em Phys.\ Rev.\
  Lett.\/} {\bf 101} 207004

\bibitem{Sefat08b}
Sefat A~S, Jin R, McGuire M~A, Sales B~C, Singh D~J and Mandrus D 2008 {\em
  Phys.\ Rev.\ Lett.\/} {\bf 101} 117004

\bibitem{Kumar08a}
Kumar M, Nicklas M, Jesche A, Caroca-Canales N, Schmitt M, Hanfland M,
  Kasinathan D, Schwarz U, Rosner H and Geibel C 2008 {\em Phys.\ Rev.\ B\/}
  {\bf 78} 184516

\bibitem{fplo1}
Koepernik K and Eschrig H 1999 {\em Phys.\ Rev.\ B\/} {\bf 59} 1743

\bibitem{fplo2}
Ophale I, Koepernik K and Eschrig H 1999 {\em Phys.\ Rev.\ B\/} {\bf 60} 14035

\bibitem{version}
We used version 7.00 of the FPLO release for all calculations pertaining to
  structural and magnetic transitions, doping and pressure effects. For the
  calculation of the electric field gradient (EFG) version 5.19 was used. As
  basis set: Ca (3$s$3$p$/4$s$4$p$3$d$+5$s$5$p$), Sr
  (3$d$4$s$4$p$/5$s$5$p$4$d$+6$s$6$p$), Ba
  (4$d$5$s$5$p$/6$s$6$p$5$d$+4$f$7$s$7$p$), Fe (3$s$3$p$/4$s$4$p$3$d$+5$s$5$p$)
  and As (3$s$3$p$3$d$/4$s$4$p$4$d$+5$s$5$p$) where chosen for
  semicore/valence+polarization states. The high lying states improve the basis
  which is especially important for the calculation of the EFG.

\bibitem{perdew}
Perdew J~P and Wang Y 1992 {\em Phys.\ Rev.\ B\/} {\bf 45} 13244

\bibitem{czyzyk}
Czy\.{z}yk M~T and Sawatzky G~A 1994 {\em Phys.\ Rev.\ B\/} {\bf 49} 14211

\bibitem{aza}
Anisimov V~I, Zaanen J and Andersen O~K 1991 {\em Phys.\ Rev.\ B\/} {\bf 44}
  943

\bibitem{laz}
Liechtenstein A~I, Anisimove V~I and Zaanen J 1995 {\em Phys.\ Rev.\ B\/} {\bf
  52} R5467

\bibitem{Morinaga2008}
Morinaga R, Matan K, Suzuki H~S and Sato T~J 2008  ArXiv:0809.3084v2

\bibitem{Bostrom2006}
Bostrom M, Prots Y and Grin Y 2006 {\em J.\ Sol.\ State Chem.\/} {\bf 179} 2472

\bibitem{Kreyssig08a}
Kreyssig A, Green M~A, Lee Y, Samolyuk G~D, Zajdel P, Lynn J~W, Bud'ko S~L,
  Torikachvili M~S, Ni N, Nandi S, Leao J, Poulton S~J, Argyriou D~N, Harmon
  B~N, Canfield P~C, McQueeney R~J and Goldman A~I 2008 {\em Phys.\ Rev.\ B\/}
  {\bf 78} 184517

\bibitem{Jesche08a}
Jesche A, {Caroca-Canales} N, Rosner H, Borrmann H, Ormeci A, Kasinathan D,
  Kaneko K, Klauss H~H, Luetkens H, Khasanov R, Amato A, Hoser A, Krellner C
  and Geibel C 2008 {\em Phys.\ Rev.\ B\/} {\bf 78} R180504

\bibitem{Tegel08a}
Tegel M, Rotter M, Weiss V, Schappacher F~M, P{\"o}ttgen R and Johrendt D 2008
  {\em J.\ Phys.: Condens.\ Matter\/} {\bf 20} 452201

\bibitem{Ma2008}
Ma C, Yang H, Tian H, Shi H, Lu J, Wang Z, Zeng L, Chen G, Wang N and Li J 2008
   ArXiv:0811.3270v2

\bibitem{Huang08}
Huang Q, Qiu Y, Bao W, Lynn J, Green M, Chen Y, Wu T, Wu G and Chen X 2008 {\em
  Phys.\ Rev.\ Lett.\/} {\bf 101} 257003

\bibitem{Wu08a}
Wu G, Chen H, Wu T, Xie Y~L, Yan Y~J, Liu R~H, Wang X~F, Ying J~J and Chen X~H
  2008 {\em J.\ Phys.: Condens.\ Matter\/} {\bf 20} 422201

\bibitem{footnote3}
Note that the structural parameters for FeSe are taken from a very old
  reference, and that plasma frequencies depend sensitively on the band
  structure, which in turn can be affected strongly by the details of the
  crystal structure.

\bibitem{Mazin08a}
Mazin I~I, Johannes M~D, Boeri L, Koepernik K and Singh D~J 2008 {\em Phys.\
  Rev.\ B\/} {\bf 78} 085104

\bibitem{footnote1}
Experimentally, the Eu spins in EuFe$_{2}$As$_{2}$ are shown to order
  anti-ferromagnetically in the $a-b$ plane. This in turn doubles the unit cell
  along the $c$-axis during the LDA+$U$ calculations and introduces to
  inequivalent As positions. Optimizing the Fe-As distance for this scenario is
  more time consuming and is therefore postponed.

\bibitem{Mazin08b}
Mazin I~I and Johannes M~D 2008  ArXiv:0807.3737v1

\bibitem{TGoko08a}
Goko T, Aczel A~A, {Baggio-Saitovitch} E, {Bud'ko} S~L, Canfield P~C, Carlo
  J~P, Chen G~F, Dai P, Hamann A~C, Hu W~Z, Kageyama H, Luke G~M, Luo J~L,
  Nachumi B, Ni N, Reznik D, {Sanchez-Candela} D~R, Savici A~T, Sikes K~J, Wang
  N~L, Wiebe C~R, Williams T~J, Yamamoto T, Yu W and Uemura Y~J 2008
  ArXiv:0808.1425v1

\bibitem{Goldman08a}
Goldman A~I {\em et~al.\/} 2008 {\em Phys.\ Rev.\ B\/} {\bf 78} R100506

\bibitem{Zhao08a}
Zhao J {\em et~al.\/} 2008 {\em Phys.\ Rev.\ B\/} {\bf 78} R140504

\bibitem{Matan08a}
Matan K, Morinaga R, Iida K,  and Sato T~J 2008  ArXiv:0810.4790

\bibitem{Yildirim08a}
Yildirim T 2008  ArXiv:0807.3936v2

\bibitem{Torikachvili08a}
Torikachvili M~S, {Bud'ko} S~L, Ni N and Canfield P~C 2008 {\em Phys.\ Rev.\
  Lett.\/} {\bf 101} 057006

\bibitem{Miclea08a}
Miclea C~F, Nicklas M, Jeevan H~S, Kasinathan D, Hossain Z, Rosner H, Gegenwart
  P, Geibel C and Steglich F 2008  ArXiv:0808.2026v1

\bibitem{Alireza08a}
Alireza P~L, Ko Y~T~C, Gillett J, Petrone C~M, Cole J~M, Lonzarich G~G and
  Sebastian S~E 2008 {\em J.\ Phys.: Condens.\ Matter\/} {\bf 21} 012208

\bibitem{Yu08a}
Yu W, Aczel A~A, Williams T~J, Bud'ko S~L, Ni N, Canfield P~C and Luke G~M 2008
   ArXiv:0811.2554v1

\bibitem{pyykko}
Pyykk\"o P 2001 {\em Mol. Phys.\/} {\bf 99} 1617

\bibitem{abragam}
Abragam A 2006 {\em {The principles of nuclear magnetism, Oxford Univ.
  Press}\/}

\bibitem{Sr122GiPa}
Tetragonal phase: $a=3.9250$, $c=12.332$, $z($As$)=0.358$, orthrhombic pase:
  $a=5.5819$, $b=5.5197$, $c=12.332$, $z($As$)=0.358$

\bibitem{Baek}
Baek S~H, Curro N~J, Klimczuk T, Bauer E~D, Ronning F and Thompson J~D 2008
  ArXiv:0808.0744v3

\bibitem{Kitagawa}
Kitagawa K, Katayama N, Ohgushi K, Yoshida M and Takigawa M 2008
  ArXiv:0807.4613v3

\bibitem{peter}
Jegli\v{c} P private communication

\bibitem{Lapaper}
Grafe H~J, Lang G, Hammerath F, Paar D, Manthey K, Koch K, Rosner H, Curro N~J,
  Behr G, Werner J, Leps N, Klingeler R and Büchner B 2008  ArXiv:0811.4508

\bibitem{Ndpaper}
Jegli\v{c} P, Bos J~W~G, Zorko A, Brunelli M, Koch K, Rosner H, Margadonna S
  and Ar\v{c}on D 2008  Submitted Phys. Rev. B

\bibitem{Saha08a}
Saha S~R, Butch N~P, Kirshenbaum K and Paglione J 2008  ArXiv:0811.3940v1

\bibitem{Park08a}
Park T, Park E, Lee H, Klimczuk T, Bauer E~D, Ronning F and Thompson J~D 2008
  {\em J.\ Phys.: Condens.\ Matter\/} {\bf 20} 322204

\bibitem{Sefat08aetal}
Sefat A~S {\em et~al.\/} 2008 {\em Phys.\ Rev.\ B\/} {\bf 78} 104505

\bibitem{CWang08aetal}
Cao G {\em et~al.\/} 2008  ArXiv:0807.1304v2

\bibitem{Tanatar08a}
Tanatar M~A, Ni N, Martin C, Gordon R~T, Kim H, Kogan V~G, Samolyuk G~D, Bud'ko
  S~L, Canfield P~C and Prozorov R 2008  ArXiv:0808.4991v1

\bibitem{LJLi08a}
Li L~J, Wang Q~B, Luo Y~K, Chen H, Tao Q, Li Y~K, Lin X, He M, Zhu Z~W, Cao G~H
  and Xu Z~A 2008  ArXiv:0809.2009v1

\bibitem{ZRen08a}
Ren Z, Tao Q, Jiang S, Feng C, Wang C, Dai J, Cao G and Xu Z 2008
  ArXiv:0811.2390v1

\bibitem{CWang08c}
Wang C, Jiang S, Tao Q, Ren Z, Li Y, Li L, Feng C, Dai J and abd Zhu'an~Xu G~C
  2008  ArXiv:0811.3925v1

\bibitem{Rotter08c}
Rotter M, Tegel M, Schellenberg I, Schappacher F~M, P\"ottgen R, Deisenhofer J,
  G\"unther A, Schrettle F, Loidl A and Johrendt D 2008  ArXiv:0812.2827v1

\bibitem{JHChu08a}
Chu J, Analytis J~G, Kucharczyk C and Fisher I~R 2008  ArXiv:0811.2463v1

\bibitem{XFWang08a}
Wang X~F, Wu T, Wu G, Liu R~H, Chen H, Xie Y~L and Chen X~H 2008
  ArXiv:0811.2920v1

\bibitem{GFChen08aetal}
Chen G~F {\em et~al.\/} 2008 {\em Chin.\ Phys.\ Lett.\/} {\bf 25} 3403

\bibitem{GMu08a}
Mu G, Luo H, Wang Z, Ren Z, Shan L, Ren C and Wen H 2008  ArXiv:0812.1188v1

\bibitem{TrisconeJunod96}
Triscone G and Junod A 1996 {\em {Bismuth-based High-Temperature
  Superconductors}\/} ed {H\ Maeda and K\ Togano} ({Marcel Dekker, New York})
  pp 33--74

\bibitem{Drechsler03aetal}
Drechsler S~L {\em et~al.\/} 2003 {\em Physica B\/} {\bf 329} 1352--4

\bibitem{Sekiba08a}
Sekiba Y, Sato T, Nakayama K, Terashima K, Richard P, Bowen J~H, Ding H, Xu
  Y~M, Li L~J, Cao G~H, Xu Z~A and Takahashi T 2008  ArXiv:0812.4111v1

\end{thebibliography}
\providecommand{\newblock}{}

\end{document}